\documentclass[12pt]{iopart}

\usepackage[backend=biber,style=numeric-comp,autocite=plain,sorting=none]{biblatex}
\usepackage{todonotes}
\usepackage{graphicx}
\usepackage{dirtytalk}
\usepackage{booktabs}
\usepackage{url}
\usepackage[maxfloats=40]{morefloats}
\usepackage{subfigure}
\usepackage[export]{adjustbox}
\usepackage{xcolor}
\usepackage[normalem]{ulem}  

\newcommand{\old}[1]{}
\newcommand{\new}[1]{#1}

\begin{document}

\title[Deep Learning-Based Multi-Event Reconstruction for Delay Line Detectors]{Deep Learning-Based Spatiotemporal Multi-Event Reconstruction for Delay Line Detectors}



\author{
Marco Knipfer$^{1,2}$,
Stefan Meier$^1$,
Tobias Volk$^1$,
Jonas Heimerl$^1$,
Peter Hommelhoff$^1$,
Sergei Gleyzer$^2$
}

\address{$^1$ Chair for Laser Physics, Department of Physics, Friedrich-Alexander-Universität Erlangen-Nürnberg (FAU),
Staudtstraße 1, 91058, Erlangen, Germany}
\address{$^2$ Department of Physics and Astronomy, University of Alabama, Tuscaloosa, AL 35487, USA}
\ead{\mailto{marco.knipfer@fau.de}}
\vspace{10pt}
\begin{indented}
\item[]\today
\end{indented}

\vspace{2pc}

\begin{abstract}
Accurate observation of two or more particles within a very narrow time window has always been a challenge in modern physics. It creates the possibility of correlation experiments, such as the ground-breaking Hanbury Brown-Twiss experiment, leading to new physical insights. For low-energy electrons, one possibility is to use a Microchannel plate with subsequent delay lines for the readout of the incident particle hits, a setup called a Delay Line Detector. The spatial and temporal coordinates of more than one particle can be fully reconstructed outside a region called the dead radius. For interesting events, where two electrons are close in space and time, the determination of the individual positions of the electrons requires elaborate peak finding algorithms. While classical methods work well with single particle hits, they fail to identify and reconstruct events caused by multiple nearby particles.
To address this challenge, we present a new spatiotemporal machine learning model to identify and reconstruct the position and time of such multi-hit particle signals. 
This model achieves a much better resolution for nearby particle hits compared to the classical approach, removing some of the artifacts and reducing the dead radius a factor of eight. We show that machine learning models can be effective in improving the spatiotemporal performance of delay line detectors.
\end{abstract}


\section{Introduction}
\begin{figure}[t]
	\centering
	\includegraphics[width=1\linewidth]{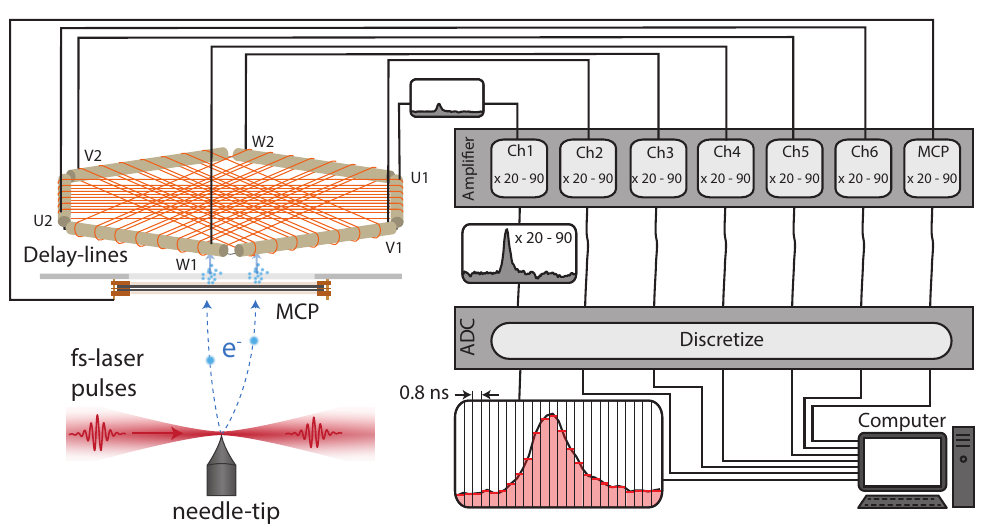}
	\caption{Experimental setup for measuring multi-electron events with a Delay Line Detector.
	Triggered by ultrashort laser pulses, the needle tip emits one or more electrons that hit the Microchannel Plate (MCP), intensified by secondary electron emission. The resulting bunch passes the delay lines and induces a voltage pulse, producing the data for 6 channels (Ch1--6) and the MCP signal.
	After amplification the data is discretized by the analog-to-digital converter (ADC).
	}
	\label{fig:setuptcorrelationchamber}
\end{figure}

The detection of single particles, such as atoms, ions, electrons and highly energetic photons is the cornerstone of many fields in fundamental physics research.
Typically, the signal due to a single particle is too faint to be measured directly and requires amplification.
It is usually achieved with electron avalanches, as for example in Photon Multiplier Tubes (PMTs)~\cite{Lubsandorzhiev2006} and Microchannel Plates (MCPs) \cite{Gys2015}.
MCPs offer the advantage that the incident particle can be spatially resolved with a phosphor screen or a wire-grid.
The latter system is called a Delay Line Detector~(DLD)~\cite{Jagutzki2002}, a spatially and temporally resolving detector, schematically shown in Figure~\ref{fig:setuptcorrelationchamber}.
DLDs can be used for ions, electrons and photons with energies high enough to emit an electron from the front side of the MCP.

In ultra-fast atomic physics, Delay Line Detectors are used for Cold Target Recoil Ion Momentum Spectroscopy (\textit{COLTRIMS})~\cite{Ullrich2003}, a technique for resolving the correlation of two simultaneously emitted electrons in non-sequential double ionization (NSDI)~\cite{Weber2000}.
In another seminal experiment, a Delay Line Detector was used to compare the bunching and anti-bunching behavior of free bosonic particles and free fermionic particles~\cite{Jeltes2007}. 
Usually the strength of correlation signals scales inversely with the separation of the particles, e.g.\ Coulomb interaction for electrons.
Therefore, it is an important task to reconstruct particles that are as close as possible without errors.
Despite the additional information provided by the three layers of delay lines, the disentangling of two or more incident particle signals becomes increasingly more challenging the closer the particles get to each other.
Although Delay Line Detectors are very effective in identifying and reconstructing single particle hits, 
they are much more limited in the ability to identify and reconstruct multi-hit events that are close in space and time.
In such scenarios, the signals will significantly overlap and present reconstruction challenges. 

Machine learning has seen an explosive increase in use in physics, including use cases for classification, regression, anomaly detection, generative modeling and others~\cite{Albertsson:2018maf, Guest:2018yhq, Larkoski:2017jix, Radovic2018-no, Carleo:2019ptp,Bourilkov:2019yoi,Schwartz:2021ftp,Karagiorgi:2021ngt,Boehnlein:2021eym}. 
Most of these studies focus exclusively on the spatial or on the temporal domain.
Recently, spatiotemporal studies and models that combine spatial and temporal input data have started gaining traction in physics and other fields~\cite{https://doi.org/10.48550/arxiv.2011.10616, 10.1145/3394486.3403198, Shi_2019, DeepGLEAM, NeuralPointProcess, CovarianceMatrixHierarchicalBayesianSpatioTemporal, SpatialModelingPrecipitation, PredictingClusteredWeatherPatterns, 2018arXiv180400684W, 2018arXiv180200386W, 2020arXiv201110616W, 10.1145/3394486.3403198}.
For a review of deep learning spatiotemporal applications, please see \cite{Karnowski2012DeepML}. Additionally, machine learning algorithms have recently been applied to the challenge of reconstructing close-by particles, for example, by the CMS Collaboration~\cite{CMS:2022wjj}.

In this work, we focus on the particularly challenging scenario where the individual signals due to multiple particles are close in space and time. We show that machine learning algorithms can significantly improve the multi-hit capability of Delay Line Detectors compared to classical reconstruction methods. Our approach can substantially reduce the effective dead radius of simultaneously arriving particles, improving the overall quality of the reconstruction. 


The content of the paper is structured as follows:
Section~\ref{sec:exp_setup} explains the experimental setup of the Delay Line Detector and data collection.
In Section~\ref{sec:classical_setup} the classical peak finding methods are described, followed by the the machine learning approach in Section~\ref{sec:ML_setup}.
In Section~\ref{sec:evaluation} we present the results, followed by their interpretation in Section~\ref{sec:interpretation}.
Section~\ref{sec:summary} provides a summary and outlook.

\section{Experimental Setup}
\label{sec:exp_setup}
The experimental setup is shown in Figure~\ref{fig:setuptcorrelationchamber}.
It consists of a tungsten needle tip with an apex radius of $r_\mathrm{tip}=5$ to $20$\,nm that is illuminated with ultra-short few-cycle laser pulses from an optical parametric amplifier system.
It has a central wavelength of $800\,$nm and a repetition rate of $f_\mathrm{rep}=200\,$kHz.
The typical pulse duration is $\tau=12\,$fs.
As the electron emission occurs promptly with the incident laser pulse, this represents an ultra-fast electron source with initial pulse durations of a few femtoseconds that are far below any electronically resolvable time scale.

The whole experiment takes place in an ultra-high vacuum chamber with a base pressure $<10^{-9}\,$mbar.
The tip is typically biased with a voltage between $U_\mathrm{tip}=-15$ to $-50\,$V that corresponds to the final energy of the emitted electrons except the influence of the laser field.
Due to the negative bias, the electrons are accelerated away from the tip and travel towards the delay line detector.
In more detail, it consists of a Chevron type (double) Microchannel plate detector.
The front MCP is a funnel-type MCP that provides a very high detection efficiency.
Each incident electron will cause secondary emission, resulting in an electron bunch of $10^5$--$10^6$ electrons.
This electron bunch is additionally accelerated towards a biased meandering wireframe called the delay line.

\begin{figure}[t]
	\centering
	\includegraphics[width=0.7\linewidth]{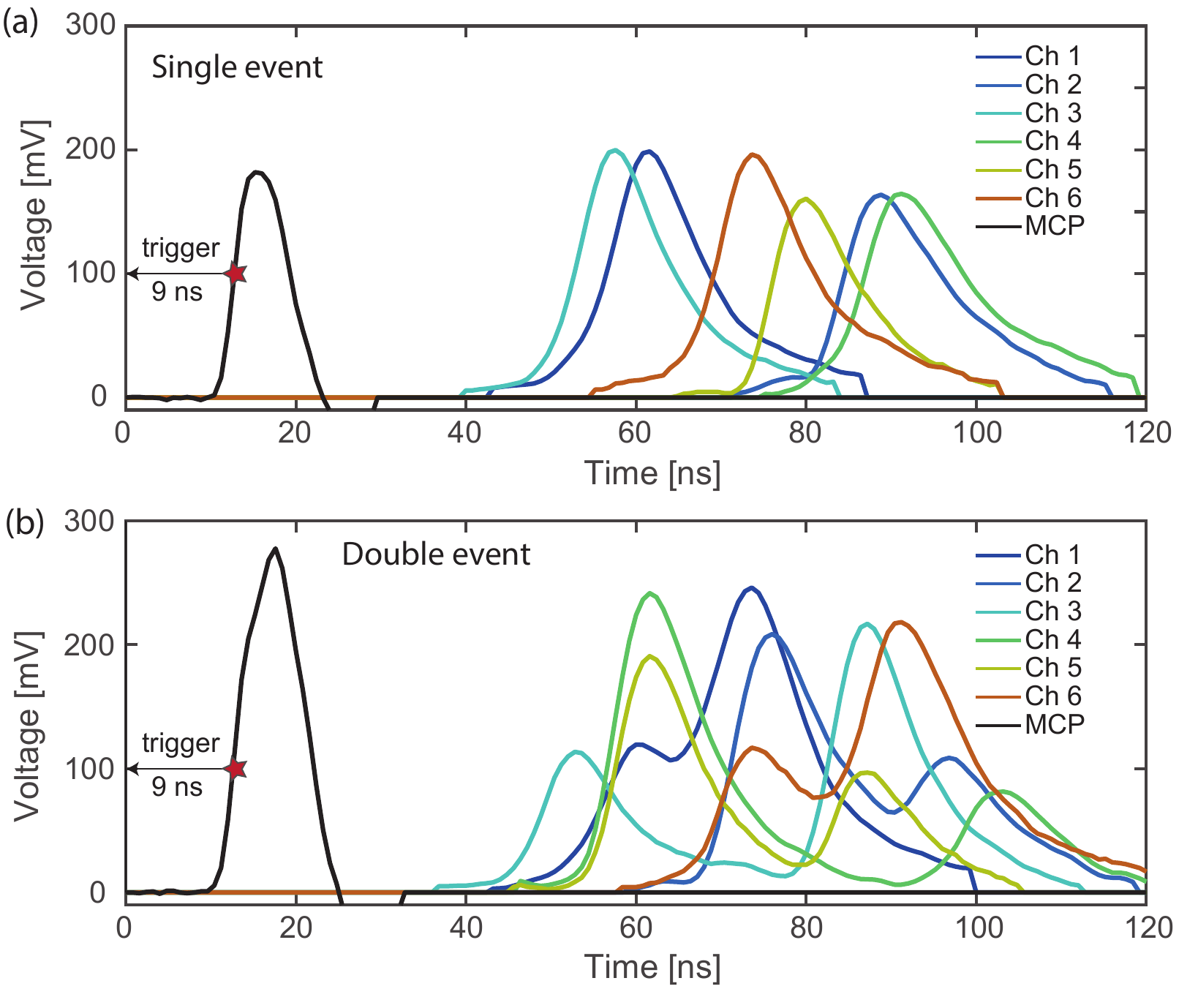}
	\caption{(a) Typical time trace of one electron hitting the Delay Line Detector.
	7 channels are recorded, consisting of 6 delay line signals and one Microchannel Plate Detector  signal.
	(b) Two-electron event, where two MCP peaks are merged into a single peak, as can be seen by the shoulder to the left.
	For each event, data recording starts 9\,ns before the trigger threshold is reached, as indicated by the arrow and the red star.}
	\label{fig:examplehit}
\end{figure}

The electron bunch will induce a voltage pulse at the same position where the incoming particle has hit the MCP.
Two voltage pulses are excited and travel in both directions in each wire of the wireframe and are detected at each end of the wire.
There are three such wireframes behind the MCP, resulting in six channels where the signals are recorded. 
Additionally, a seventh voltage signal can be picked up from the supply voltage of the MCP with the help of a bias tee.
In total, the detector output counts up to 7 signals that are fed out of the vacuum chamber, amplified and digitized by a fast analog-to-digital converter (ADC) with a bin-size of $0.8\,$ns.
Typical single and double hit event time traces are shown in Figure~\ref{fig:examplehit}.

\section{Spatiotemporal Signal Reconstruction Methods}   
\label{sec:classical_setup}
The classical approach for analysis of Delay Line Detector reconstruction data consists of two commonly used methods:
the hardware-based Constant Fraction Discriminator (CFD) and the Fit-based Method (FM).
\subsection{Hardware-Based Time Stamp Detection}
\begin{figure}
    \centering
    \includegraphics[width=1\linewidth]{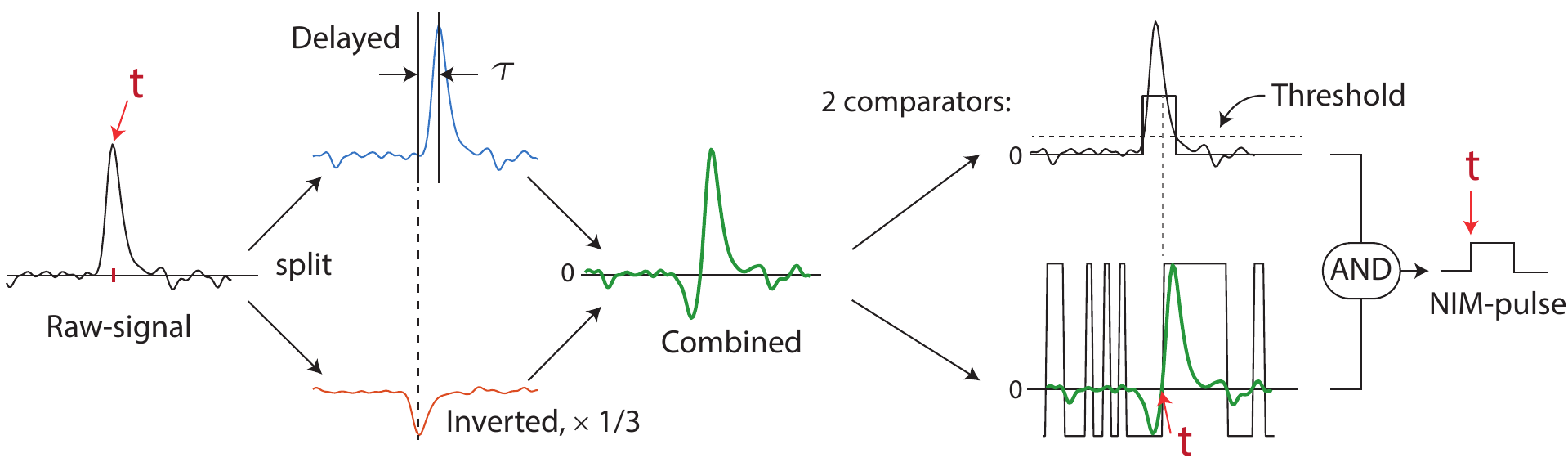}
    \caption{Principle of the constant fraction discriminator (CFD): An incoming pulse is split into two parts. The upper one is delayed, the lower one is inverted and multiplied by a factor $<1$. Both pulses are combined and sent into two comparator circuits. A NIM pulse is created whose rising edge contains the time information.}
    \label{fig:CFD}
\end{figure}

For commercial Delay Line Detectors, the determination of the $(x,\,y,\,t)$ information for the incoming particles relies on a hardware-based detection of the time stamps from the anodes and the MCP, combined with an evaluation software. 
In commercial DLDs, the signals are fed into a Constant Fraction Discriminator (CFD) after amplification. The CFD principle works as shown in Figure~\ref{fig:CFD}: an incoming signal is superimposed with its copy that was electrically reversed and reduced by a factor of about 3. Additionally, the original signal was shifted in time, which results in a well-defined zero crossing that defines the point in time where that specific signal arrived. To determine the zero crossing purely electrically, two comparators are used: (1) one that gives a logic level one when a certain threshold is exceeded, while (2) the second one gives a logic one when the signal is above the zero-voltage level (noise-level). The time-stamp is given by the point where both comparators are one. The advantage of this method is that it is independent of the actual pulse height.
The CFD outputs a Nuclear Instrument Module (NIM) pulse with a well-defined shape whose rising flank is at the temporal position of the peak of the original signal.
This NIM pulse is detected by a time-to-digital converter that digitizes the timing information with a resolution of \(\sim25\,\)ps.
We will refer to this approach as the \textit{CFD method}. 

\subsection{Multi-Hit Capability of the Hardware-Based Time Stamps}
\begin{figure}
    \centering
    \includegraphics[width=0.9\linewidth]{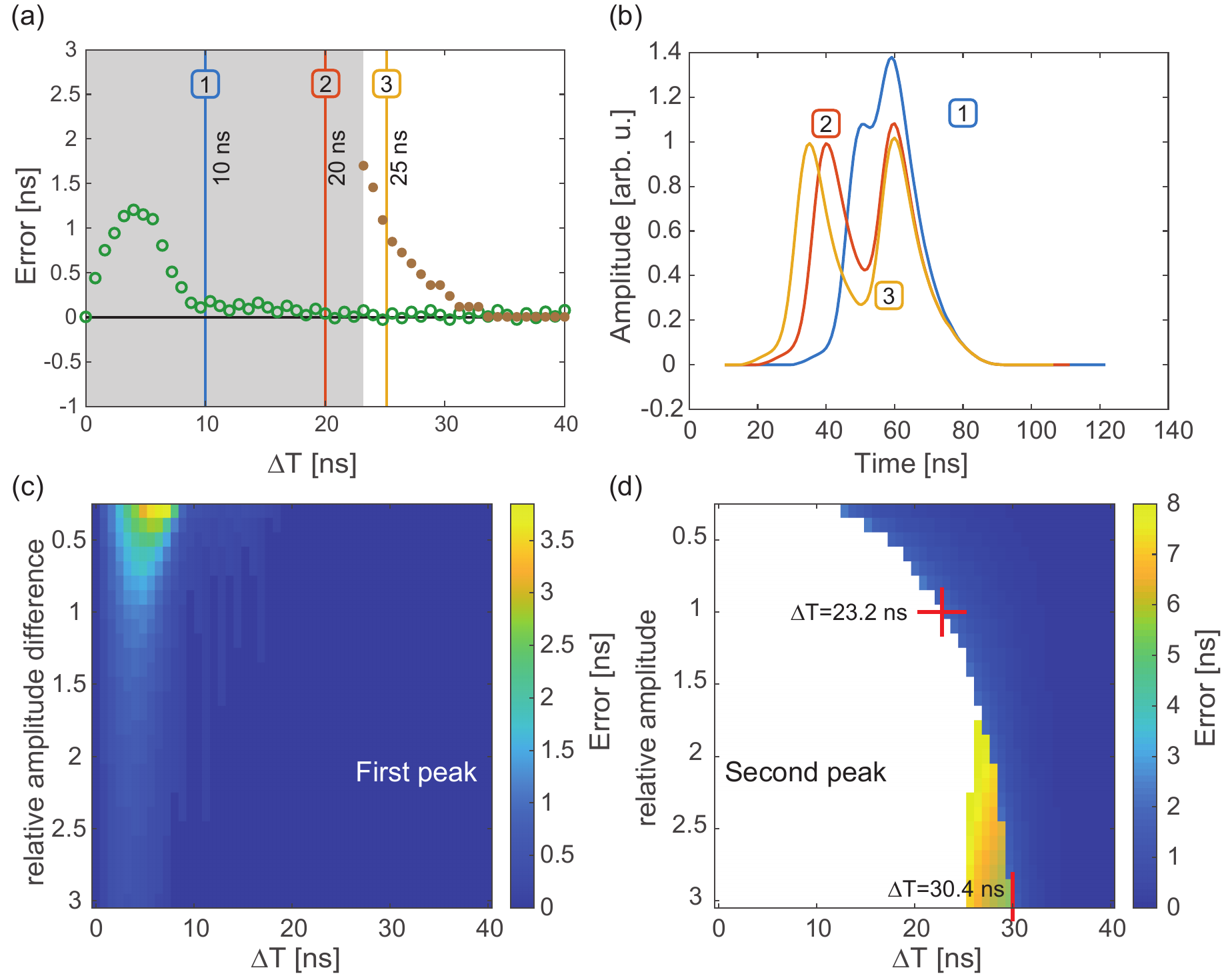}
    \caption{(a) Error of Constant Fraction Discriminator (CFD)-based double-hit evaluation.
    Below \(\Delta T\approx23\)\,ns only one peak position can be found. The green data show the error of the first peak, the brown of the second.
    The curve was created by application of the CFD principle to a norm pulse in a simulation.
    The gray area shows the time differences where the CFD only finds one peak.
    (b) Different examples of double hits, created by the sum of two norm pulses with a time delay of 10\,ns (blue, 1), 20\,ns (orange, 2) and 25\,ns(yellow, 3).
    Their respective governed peak positions are shown in (a).
    In both (a) and (b) the amplitudes of the two peaks were equal.
    (c) Error of the first peak as function of relative peak height difference and time difference \(\Delta t\).
    (d) Same map as (c) but now for the second peak.
    The white area means that no second peak was found.}
    \label{fig:CFDdoublehit}
\end{figure}

The time tags produced by the CFD method are robust, fast and easy to handle.
However, their multi-hit capability is limited when it comes to close-by events.
As shown in Figure~\ref{fig:CFDdoublehit}, two equal peaks will result in only one zero crossing as soon as they are closer than 1.9 times their Full Width at Half Maximum (FWHM). In absolute numbers, as the signals propagate with an average speed of 0.9\,mm/ns, this leads to an average dead radius of 21\,mm, as the FWHM of the norm-pulse\footnote{\new{For the norm pulse, we shifted $10^4$ signals to the same peak position and calculated their mean.}} is $\sim 12.5\,$ns.
Even before the dead radius is reached, errors on the reconstructed peak positions occur, as shown in the white area in Figure~\ref{fig:CFDdoublehit}~(a).
If two particles hit the detector with a spatial distance smaller than the dead radius, only one particle will be reconstructed.
Its position will also be subject to error, as shown in Figure~\ref{fig:CFDdoublehit}~(c).
The second particle will have even larger errors, as can be seen in Figure~\ref{fig:CFDdoublehit}~(d).
The relative amplitude difference of the two peaks matters.
If the distance between two particles is smaller than the dead radius, this shift is even more pronounced, and will result in one registered particle, as indicated by the  white area in Figure~\ref{fig:CFDdoublehit}~(d) where no detection possible.

\subsection{Analog Data Readout for Improved Classical Model Evaluation}
\label{sec:fit-based-method}
For applications that require better multi-hit capability compared to the basic hardware-based method, it is possible to purchase and use fast Analog-to-Digital Converter (ADC) cards to digitize each analog trace and use computer-based algorithms to obtain the timing information.
The ADC cards have a data rate of 300\,MB/s providing a theoretical measurement speed of \(\sim400\)\,kHz, assuming an average data length of \(\sim50\)\,bins.
After the data is recorded, we perform an offline data evaluation following the measurements 
using \textsc{MATLAB}~\cite{MATLAB:2018} and Python~\cite{van1995python} to analyze the events.
The software of the ADC cards will split multiple events on a single channel if they are separated long enough that the amplitude is lower than a pre-set trigger threshold.
In this case, we first run a routine that interpolates all parts of each channel to the same time grid.
In the next step, we use a peak finding algorithm in \textsc{MATLAB}. 
This classical algorithm searches for the indices of peak maxima
and performs a fit on a window around them. 
\begin{figure}
    \centering
    \includegraphics[width=0.9\linewidth]{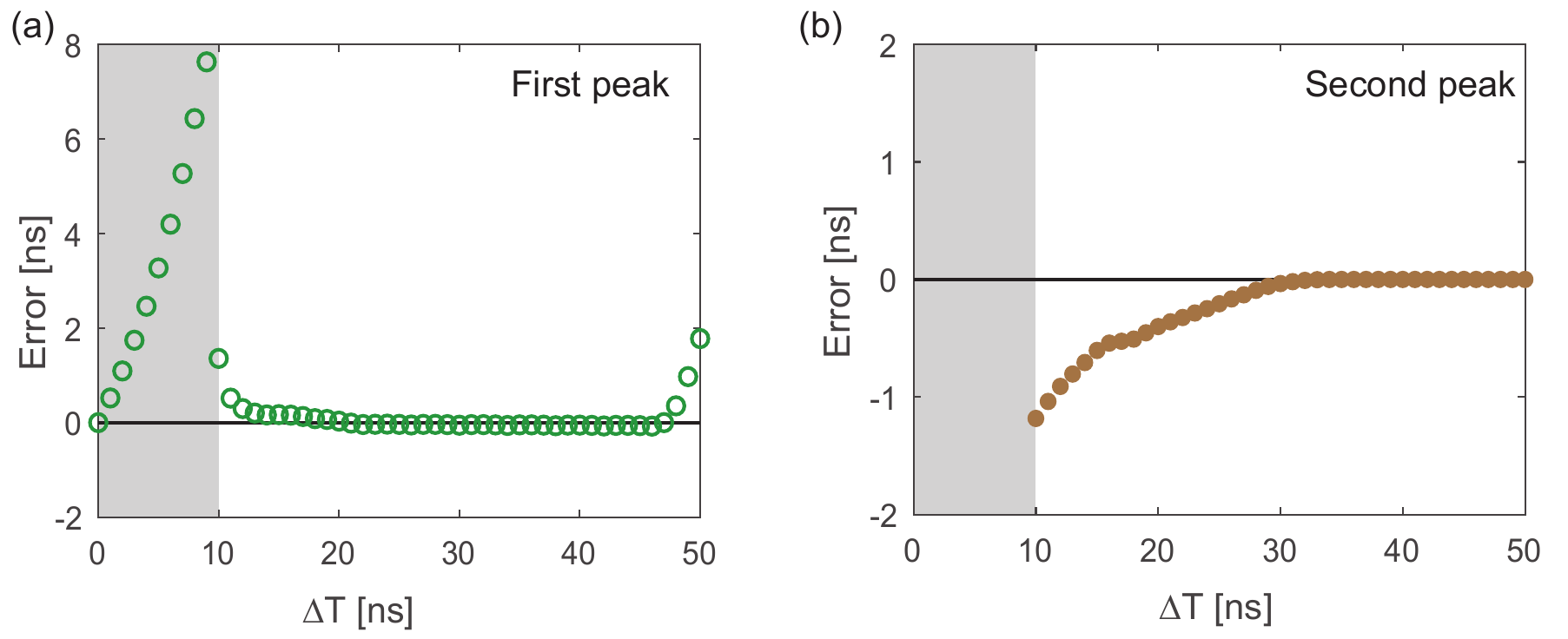}
    \caption{(a) Absolute error on the first peak for a constructed double hit using two norm pulses evaluated with the fit-based classical algorithm.
    (b) Absolute error on the second peak by the fit-based classical algorithm.}
    \label{fig:analog_error}
\end{figure}
In Figures~\ref{fig:analog_error}~(a) and~(b), the classical algorithm is used to generate the first and second peaks, with the graphs displaying the errors of these peaks as a function of their separation.
We tried different functions and selected the skewed normal distribution due to its superior performance.
The fit window is 5 bins to the right and left.
If another peak is closer than 20 bins, the window is shrunk to 2 on each side and the fit function is changed to a quadratic function.
The peak positions are passed to a position calculation software that reconstructs the event and finally calculates the position (x, y, t). 
We call this approach \textit{fit-based classical method}. 





\subsection{Mean Pulse Curve Fit}
\begin{figure}[tb]
    \subfigure[Mean pulses of different channels.]{
        \includegraphics[width=0.49\textwidth]{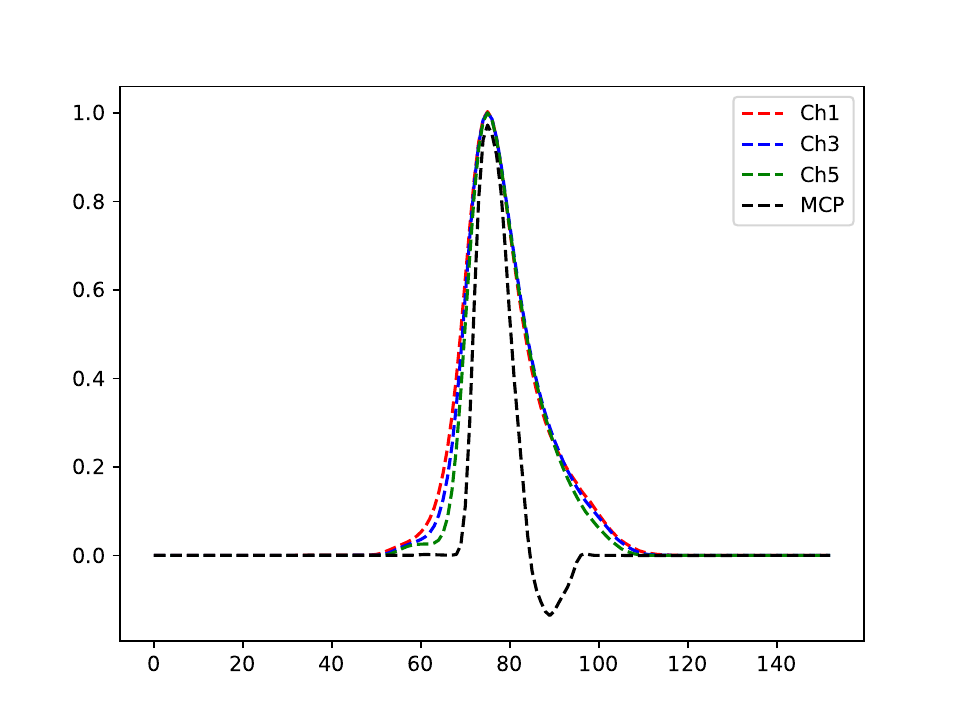}
        \label{fig:channels_comparison}
    }
    \hfill
    \subfigure[Mean puls from channels 1--6 (red) and all signals that went into the mean (gray). ]{
        \includegraphics[width=0.49\textwidth]{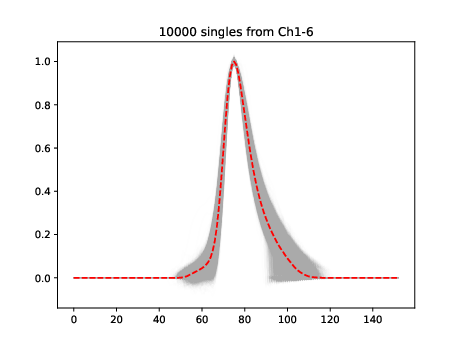}
        \label{fig:MPF}
    }
    \caption{Calculation of the mean pulse. Because the mean pulses of channels 1--6 are so similar (a), we combined them to one mean pulse (b).}
    \label{fig:MPF_construction}
\end{figure}

This method is motivated by~\cite{Wallauer2011,Lin2015, bauer:Thesis}.
Shifting and scaling single hit signals to the same position and height allows the extraction of a \say{mean pulse}.
One can then fit the double peak signal to a sum of two mean pulses to reconstruct the two peaks.
We call this method the \say{mean pulse curve fit} (MPCF).
The mean pulse can be seen in Figure~\ref{fig:MPF_construction}.
In Figure~\ref{fig:channels_comparison} it is shown that Channels 1--6 have a very similar mean pulse, only the MCP mean pulse has a different form.
Thus, we combined channels 1--6.
Figure~\ref{fig:MPF} shows 10k signals from channels 1--6 (gray) and their mean (red).
Each signal consists of a fixed number of points, in our case this is 153 time bins.
In order to be able to shift the mean pulse on a sub-bin level, the mean pulse is interpolated.
The fit function then consists of a sum of two interpolated mean pulses, each shifted, scaled and stretched.
There are six fit parameters, $[x_0, x_1, h_0, h_1, s_0, s_1]$, the peak positions $x_i$, the peak heights $h_i$ and the peak stretches $s_i$, $i\in \{0, 1\}$.
Using the \verb!curve_fit! function in \verb!scipy.optmize!~\cite{2020SciPy-NMeth} the six fit parameters are determined.
The fit is initialized with random parameters.

The quality of the fit depends upon the initial conditions and often the outcome is visibly wrong.
Several improvements are needed.
First, bounds are added to the fit parameters.
For the positions $x_i$ the bounds are the x-interval, in this case $[0,153]$.
Further we restrict the heights $h_i$ to $[0.15 h_\mathrm{max}, 1.25 h_\mathrm{max}]$, where $h_\mathrm{max}$ is the maximal value of the given signal.
Limiting the amplitude, especially in scenarios with closely spaced peaks, is beneficial.
This is because a single average pulse can often provide a good fit, leading to the second peak's amplitude going to zero.
With similar reasoning, the stretch is constrained to $s_i = \pm 70\%$.
Additionally, upon executing the fit using the mentioned scipy function, we assess the root mean square error (RMSE). If this RMSE surpasses a predefined threshold, $\mathrm{RMSE}_{\mathrm{max}}$, we reinitiate the fitting process with a new set of randomly chosen initial parameters.
$\mathrm{RMSE}_\mathrm{max}$ is initially set at $3\%$ of $h_{\mathrm{max}}$.
Following 13 unsuccessful attempts, the threshold $\mathrm{RMSE}_{\mathrm{max}} $ is incrementally increased by a factor of $1.15$ each time the fit failed to be better than the current $\mathrm{RMSE}_\mathrm{max}$.


A better way of initializing the fit parameters is to perform a fast peak finding algorithm on the signal.
We use the \verb!argrelextrema! function in scipy to get the positions of the relative maxima in the signal.
Since the fit often does not fulfill the $\mathrm{RMSE}_\mathrm{max}$ bound and to have more attempts with slightly different initial conditions, we take the initial positions $x_i$ by sampling from a normal distribution with the outputs of \verb!argrelextrema! as the means and a variance $\sigma$ that increases with every attempted fit, starting with $\sigma=20$.
The initial heights $h_i$ are sampled from a normal distribution with mean $0.75 h_\mathrm{max}$ and variance $\sigma=h_\mathrm{max}/5$.
The initial stretches $s_i$ are also randomly sampled from a normal distribution around 1.0 with variance 0.3.
Lastly, after evaluation of the MPCF method on simulated double peak signals, we found that the MPCF had a general bias in one direction.
We find the peak positions of 10k simulated doubles with the MPCF and calculate the mean error by comparison with the true peak positions.
In our case the bias was -0.53, meaning $y_\mathrm{pred} - y_\mathrm{true}$ was on average -0.53. To compensate, we subtract this bias from all predictions made by the MPCF method.

\begin{figure}
    \centering
    \includegraphics[width=\textwidth]{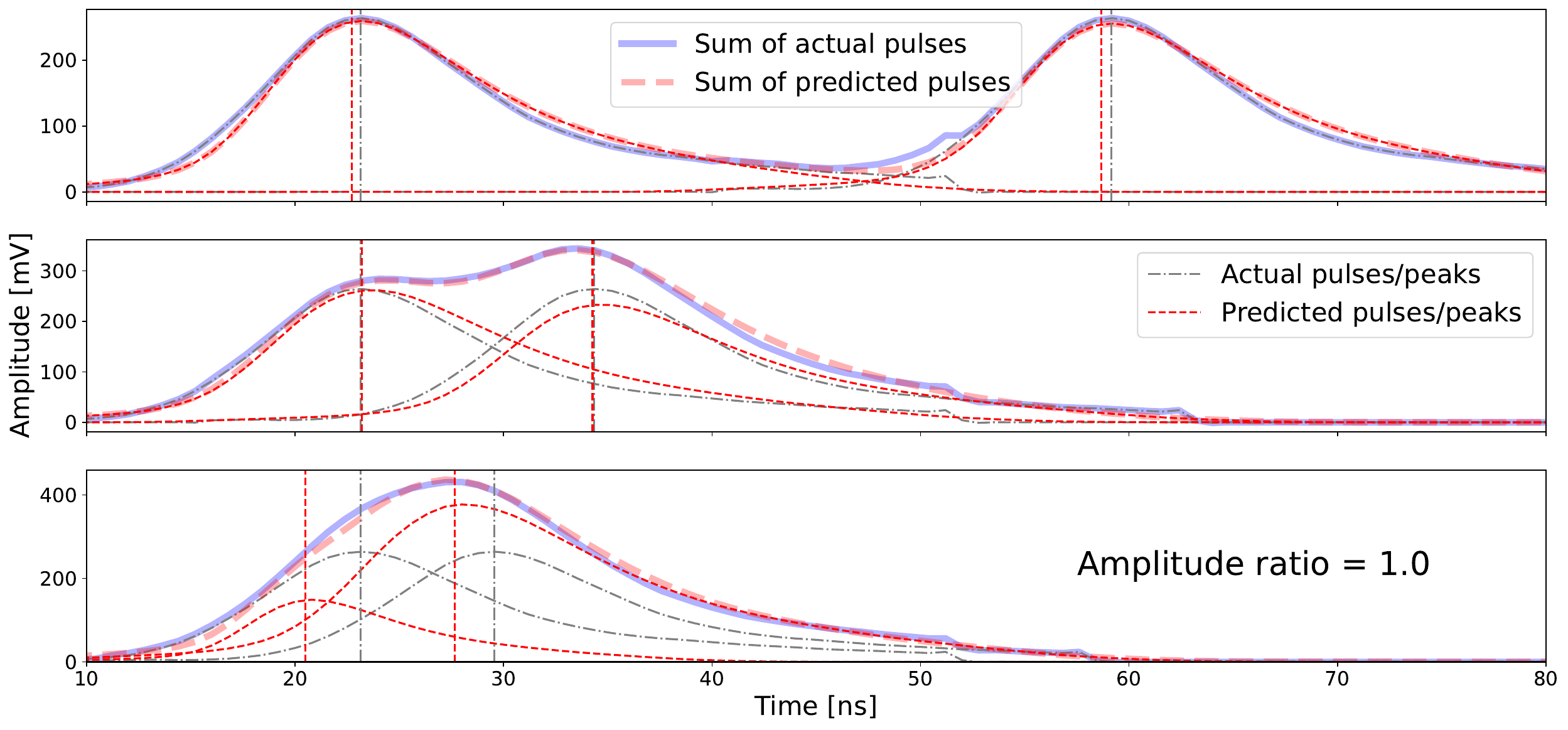}
    \caption{Three examples of predictions from the mean pulse fit method. Top: Far apart peaks, middle: close, but still two peaks visible, bottom: close and only one peak visible.
    Gray: actual pulses/peaks, red: predicted pulses/peaks.
    The actual pulses all have the same amplitude.
    }
    \label{fig:mpcf_examples}
\end{figure}

\new{Examples of predictions using the MPCF method are shown in Figure~\ref{fig:mpcf_examples}.
The MPCF is performed on data simulated from two random real single hit signals.
The mean pulse used for fitting does not exactly match the shapes used for generation.
Thus, for closer signals, the MPCF starts deviating from the ground truth, see the bottom plot in Figure~\ref{fig:mpcf_examples}.
}

\subsection{Machine Learning Approach}
\label{sec:ML_setup}
\begin{figure}[t] \label{fig:workflow}

    \centering
    \subfigure[Simulation and training pipeline]{
        \includegraphics[width=0.9\textwidth, trim=0cm 22cm 0cm 0cm, clip]{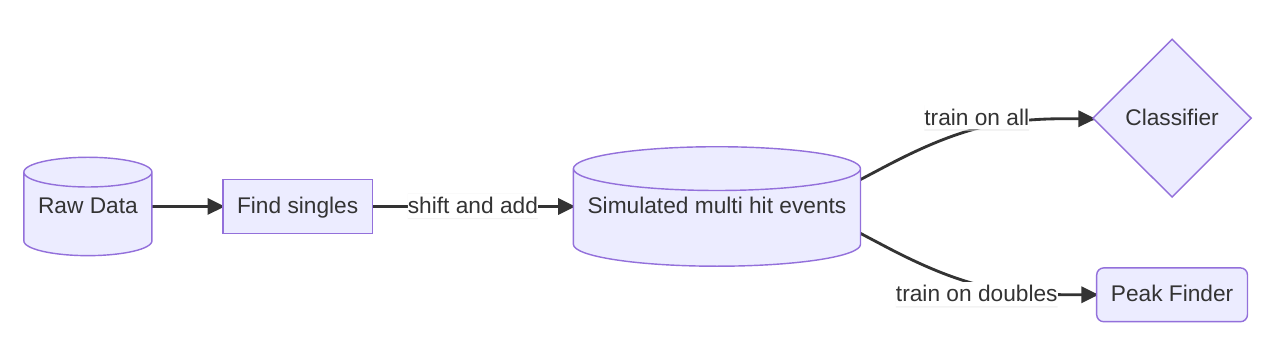}
        \label{fig:models_overview_training}
    }
    \subfigure[Inference pipeline]{
        \includegraphics[width=0.9\textwidth, trim=0cm 25cm 0cm 0.85cm, clip]{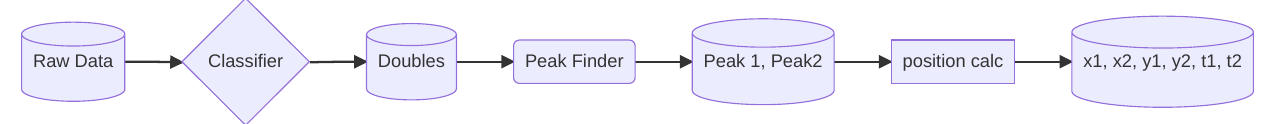}
        \label{fig:models_overview_inference}
    }
    \caption{
    (a) Simulation and training pipeline: real singles are shifted and added to create simulated multi hit events for classifier and peak finder training (b) inference pipeline: the classifier 
    determines if the event has one, two, three or four hits. For the double hit events, every channel is fed separately into the Deep Peak Finder model.
    }
    \label{fig:models_overview}
\end{figure}

The machine learning approach that we use to identify and reconstruct multi-hit events for Delay Line Detectors consists of \textit{multi-hit data generation}, a \textit{Hit Multiplicity Classifier} (HMC) and \textit{Deep Double Peak Finder} (DDPF). It is summarized in Figure~\ref{fig:models_overview}. 


\paragraph{\textit{Simulation and training workflow:}} The raw analog data is preprocessed for each event, including analog to digital conversion and zero-padding to ensure a constant vector length. From real single particle events, we generate multi-particle hit events, including doubles, triples and quadruples and use them to train a Hit Multiplicity Classifier (HMC) to distinguish among the different hit multiplicity events.
Next, we train a Deep Double Peak Finder (DDPF) model to predict the peak positions for double events, as shown in Figure~\ref{fig:models_overview_training}.

\paragraph{\textit{Inference workflow:}} After the data is pre-processed,
we apply the HMC model to classify the events into singles, doubles, triples and quadruples by using 6 out of the 7 channels (excluding the MCP channel).
Next, we apply the DDPF model to the classified events to identify the peak positions for each channel.
These peak positions are passed to a position calculation software that matches the peaks and reconstructs the event.
The final outputs are the $x$ and $y$ positions, as well as the time $t$ for each particle as shown in Figure~\ref{fig:models_overview_inference}. Our approach is general, and can be applied to any hit multiplicity once the classifier is applied. We restrict ourselves to the analysis of double-hit events for a detailed evaluation and comparison with classical methods and leave further studies of higher multiplicity events to future work. For single hit events classical algorithms work well, and no machine learning models are necessary. 

\subsubsection{Multi-Hit Data Generation}
\label{subsec:datagen}
\begin{figure}[t]
    \centering
    \includegraphics[width=0.9\textwidth]{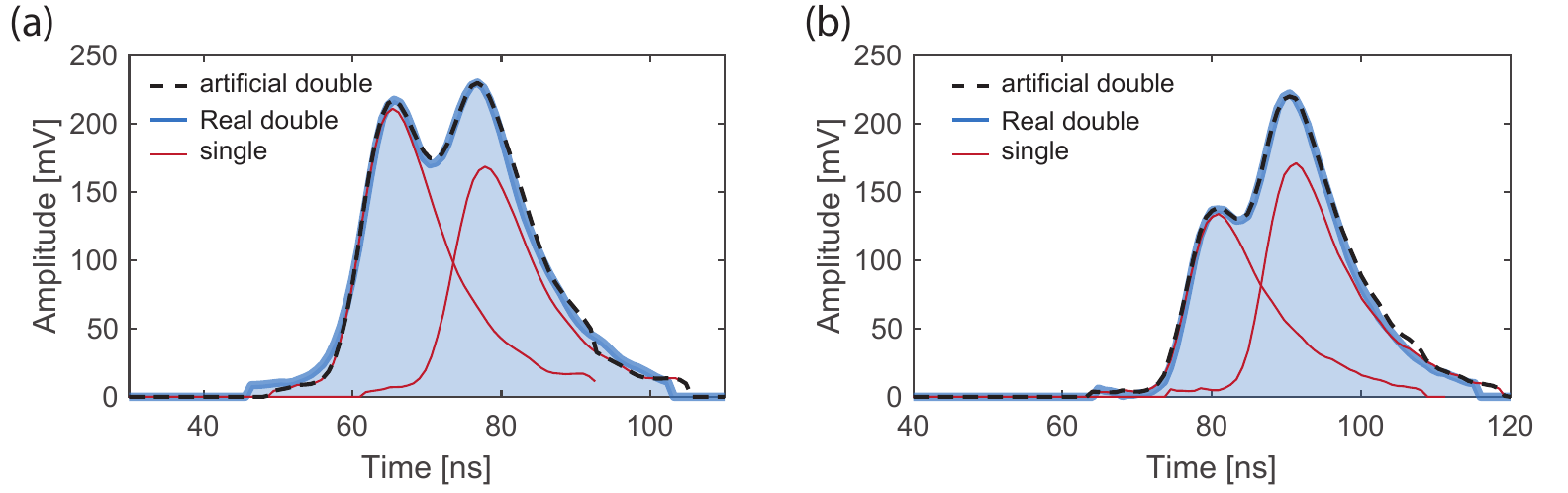}
    \caption{Examples of real double hits in different channels compared to constructed simulated double hits.
    Blue: real measured double hit, red: single hits that have been shifted and scaled so that their sum corresponds to the real double hit.
    Black dashed line: simulated double hit from the addition of the singles.
    }
    \label{fig:artificial_real_double_comparison}
\end{figure}

~\\
As the voltage signals of events are approximately additive, as can be seen in Figure~\ref{fig:artificial_real_double_comparison}, it is possible to simulate multi-hit events through addition of single peak signals to obtain doubles, triples and quadruples.
The advantage of this approach is the exact knowledge of the individual peak positions without any shifts caused by a secondary hit.
To obtain the single peak events needed for the simulated multi-hit events, we use the following filtering strategy:
\begin{itemize}
\item  We first employ the classical peak finder algorithm. Events with multiple peaks in any channel are excluded.
\item Each event is normalized by dividing the values in each channel by the event's maximum value. We then aggregate the values across all channels for each event to compute a unique metric, termed the \say{event area}. Events exceeding a threshold of the mean area plus a constant factor are discarded. They are probably double or triple hit events which could not be distinguished by the classical algorithm.
\item Using the remaining single hit candidates, doubles, triples and quadruples are simulated and an intermediate classifier model is trained on them. This classifier is then employed to further refine the remaining events.
\end{itemize}
Although this process can be iterated, we observe that additional iterations beyond the first do not significantly alter the remaining events.
To create data for the Hit Multiplicity Classifier (HMC), random events undergo simultaneous shifts in time in all channels (with one random shift for each event) and are then added across all channels concurrently.
To simulate double-hit signals for the Deep Double Peak Finder (DDPF), two single-hit signals are randomly shifted in time and added.
The shifting ensures that the model also sees peaks in regions where single hit signals might not show peaks, but in actual double hit signals, peaks would be present. 
The peak positions in the generated events are known up to the same accuracy as for the single-hit events.
Since the peak finder model predicts the peaks for each channel separately, we simulate double events for each channel separately and then shuffle the data of the different channels.


Figure~\ref{fig:artificial_real_double_comparison} shows two typical real and simulated double peaks.
The real and simulated doubles look almost identical except at the edges of the signals that are far away from the center of the peaks.
These parts of the signals are less relevant as they do not contain additional timing information of the peaks.



\subsubsection{Hit Multiplicity Classifier (HMC)}
~\\
\begin{figure}[t]
    \centering
    \includegraphics[width=0.5\textwidth]{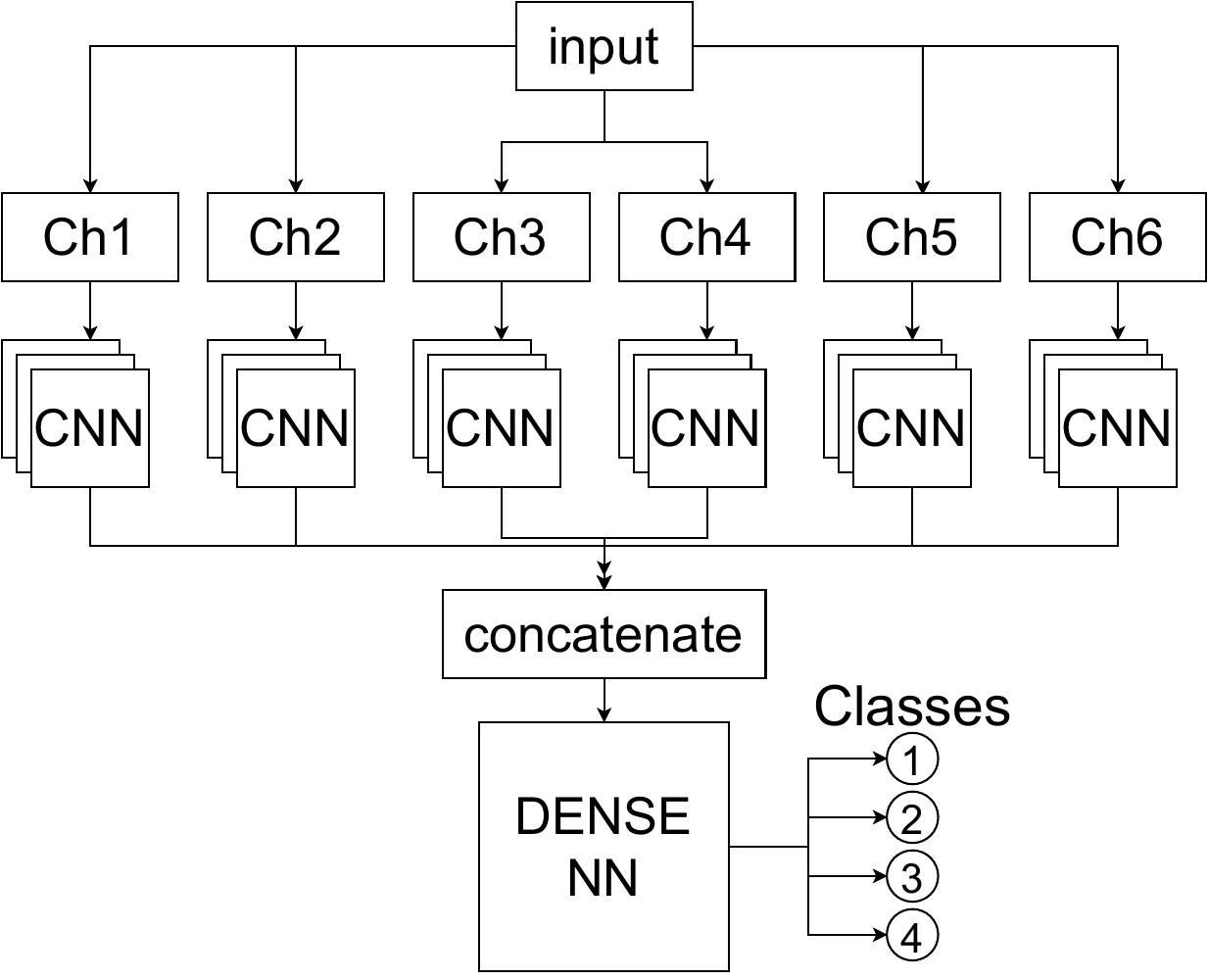}
    \caption{Hit Multiplicity Classifier model structure: The input is split into the channels Ch1--6 (without MCP) and each channel separately goes through a 1D CNN. 
    The outputs of the CNNs are concatenated and put through a simple dense NN. 
    The output of the dense NN are the probabilities for the classes [1,2,3,4].}
    \label{fig:classifier_model}
\end{figure}
A sketch of the Hit Multiplicity Classifier (HMC) is shown in Figure~\ref{fig:classifier_model}.
The input amplitudes are normalized on an event basis and split into 6 channels. Each channel is passed to a 1D convolutional neural network (CNN) that consists of: [Conv1D, BatchNorm, ReLu, Conv1D, BatchNorm, ReLu, GlobalAveragePooling] layers.
The hyperparameters \verb!num_filters! and \verb!kernel_size! are optimized during training.
After flattening, the outputs of the CNNs get concatenated and go through a dense neural network with [Dense, Dropout, Dense, Dropout] layers, where the number of neurons, dropout and activation are hyperparameters.
The last layer has 4 neurons, one for each class, with softmax activation outputs.



\subsubsection{Deep Double Peak Finder (DDPF)}
\label{sec:pf_setup}
\begin{figure}[t]
    \centering
    \includegraphics[width=\textwidth]{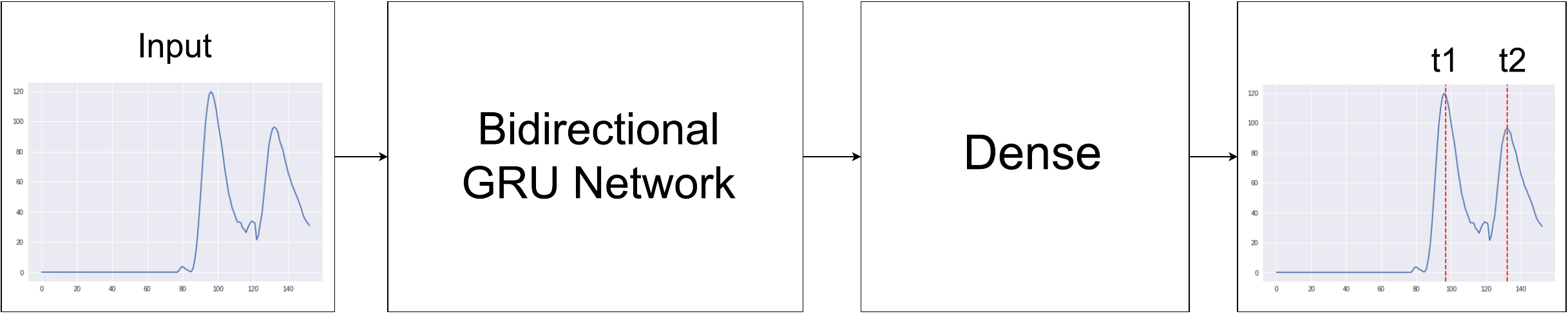}
    \caption{Deep Peak Finder model structure: The model operates on single channel data.
    After a bidirectional GRU network, a dense network predicts the two peak positions.
    }
    \label{fig:peak_finder_model_structure}
\end{figure}

~\\
A sketch of the Deep Double Peak Finder (DDPF) model for double-hit events is shown in Figure~\ref{fig:peak_finder_model_structure}.
Using data from one of the six channels at a time (without MCP), the model predicts two peak positions.
The signal amplitude is normalized for each signal and passed into two bidirectional GRU (Gated Recurrent Unit) layers~\cite{2014GRU}.
If the second GRU layer returns sequences, its output is flattened.
A GRU is very similar to a long short-term memory (LSTM) network~\cite{LSTM} cell with a forget gate.
A sketch of a GRU can be seen in Figure~\ref{fig:peak_finder_model_structure}.
We use a bidirectional GRU, i.e.\ the model steps through the time series in both directions.
Next, a set of [dropout, dense] layers is used.
A dense layer with 2 output neurons with sigmoid activation, one for each peak, produces the output of the network.
We use the sigmoid activation function because the peak positions are scaled to [0,1]. 


\subsubsection{MCP Deep Double Peak Finder}
~\\
The MCP signals have a slightly different shape and a reduced signal width compared to the other channels. Therefore, a separate model is trained for the MCP. Single events are shifted and added to create simulated doubles.  The MCP model has the same structure that is described above. The models hyperparameters are tuned during training on simulated MCP doubles.

\subsubsection{Training Protocol}
~\\
For each model, we performed hyperparameter tuning on a smaller dataset using Keras Tuner~\cite{omalley2019kerastuner} and the Hyperband optimizer~\cite{2016Hyperband}. 
Models with the best hyperparameters were retrained on the full dataset. 

The HMC model is trained with categorical cross-entropy as the loss function. 
The optimal hyperparameters were:  
\verb!activation!: 'relu', \verb!dense_1!: 132, \verb!dense_2!: 144, \verb!dropout!: 0.4, \verb!kernel_size_1!: 17, \verb!kernel_size_2!: 17, \verb!learning_rate!: 0.01, \verb!num_filters!: 55. After hyperparameter optimization, we retrained the best model for 300 epochs on 400k events (100k per class) using early stopping, keeping the best model based on validation loss and reduction of learning rate on plateau. 



For the DDPF model for channels 1--6, we use mean square error (MSE) as the loss function.
The optimal hyperparameters were: \verb!GRU_units!: 70, \verb!activation!: selu, \verb!clipnorm!: 0.001, \verb!dropout!: 0.2, \verb!hidden_layer_dim!: 100, \verb!learning_rate!: 0.001, \verb!num_layers!: 1, \verb!ret_seq!: 0.
The best model was then retrained on 1M simulated doubles for 300 epochs with a batch size of 128.

The MCP DDPF model has the same setup and MSE as the loss function.
The optimal hyperparameters were: 
\verb!GRU_units!: 65, \verb!activation!: selu, \verb!clipnorm!: 0.001, \verb!dropout!: 0.2, \verb!hidden_layer_dim!: 170, \verb!learning_rate!: 0.001, \verb!num_layers!: 1, \verb!ret_seq!: 0.

\section{Results}
\label{sec:evaluation}

\subsection{Hit Multiplicity Classifier Results}
The Hit Multiplicity Classifier model has a test accuracy of 0.9973.
for the test data consisting of 1M events evenly split into singles/doubles/triples/quadruples.
The area under the ROC curve (AUC) is $>0.9998$ for every class (one-vs-all) and thus also for the macro average.
\begin{figure}[t]
    \addtolength{\tabcolsep}{-3pt}  
    \centering
    \subfigure[]{
    \adjustbox{valign=c}{
    \begin{tabular}{@{}cllll@{}}
    \toprule
    \multicolumn{1}{l}{t\textbackslash{}p} &  1 & 2 & 3 & 4 \\ \midrule
    1                                                 & 249,954   & 46  & 0  & 0  \\
    2                                                 & 102   & 249,787  & 110  & 1  \\
    3                                                 & 0   & 340  & 249,235  & 425  \\
    4                                                 &  0  & 4 & 1670  &  248,326 \\ \bottomrule
    \end{tabular}
    }
    }
    \subfigure[]{
    \adjustbox{valign=c}{
    \begin{tabular}{@{}cllll@{}}
    \toprule
    \multicolumn{1}{l}{\#} &  1 & 2 & 3 & 4 \\ \midrule
      1 & \underline{9.9e-01}& 3.2e-06& 4.1e-18& 4.7e-22 \\
      2 & 0.0e+00& 0.0e+00& 3.8e-24& \underline{1.0e+00} \\
      3 & 2.4e-23& 2.4e-18& \underline{9.9e-01}& 8.3e-05 \\
      4 & 1.8e-17& \underline{1.0e+00}& 2.1e-11& 1.1e-18 \\
      5 & 1.1e-17& \underline{1.0e+00}& 4.3e-10& 1.3e-16 \\
      6 & 0.0e+00& 0.0e+00& 1.3e-34& \underline{1.0e+00} \\
      7 & \underline{9.9e-01}& 2.1e-06& 3.4e-14& 1.5e-18 \\
    \bottomrule
    \end{tabular}
    }
    }
    \caption{(a) Confusion matrix (true$\backslash$predicted) for the Hit Multiplicity Classifier tested with 1,000,000 evenly split test events.
    (b) Hit Multiplicity Classifier probabilities (trunkated at 2 digits) for some random events.
    The most probable event class is underlined.
    All given events were predicted correctly. As can be seen from the confusion matrix, the classifier is very accurate, which results in the very high area under the curve (AUC) of $>0.9998$ (one-vs-all).
    }
    \addtolength{\tabcolsep}{2pt}  
\end{figure}

\subsection{Deep Double Peak Finder Results (Channels 1--6)}
\label{subsec:peak_finder_results}
On the training data, the root mean square error (RMSE) is $\mathrm{rmse}_\mathrm{train}=0.77~\mathrm{ns}$ and the mean average error (MAE) is $\mathrm{mae}_\mathrm{train} = 0.20~\mathrm{ns}$. 
Evaluated on 6M simulated double-hit events, the final model has an RMSE for the peak position $\mathrm{rmse}_\mathrm{test} =  0.82~\mathrm{ns}$ and the MAE is $\mathrm{mae}_\mathrm{test} =  0.20$. The bin size of the data is 0.8~ns.

\begin{figure}[t!]
    \centering
    \subfigure[]{
    \includegraphics[width=0.46\textwidth,  trim={2cm, 0.5cm, 2cm, 0.5cm}]{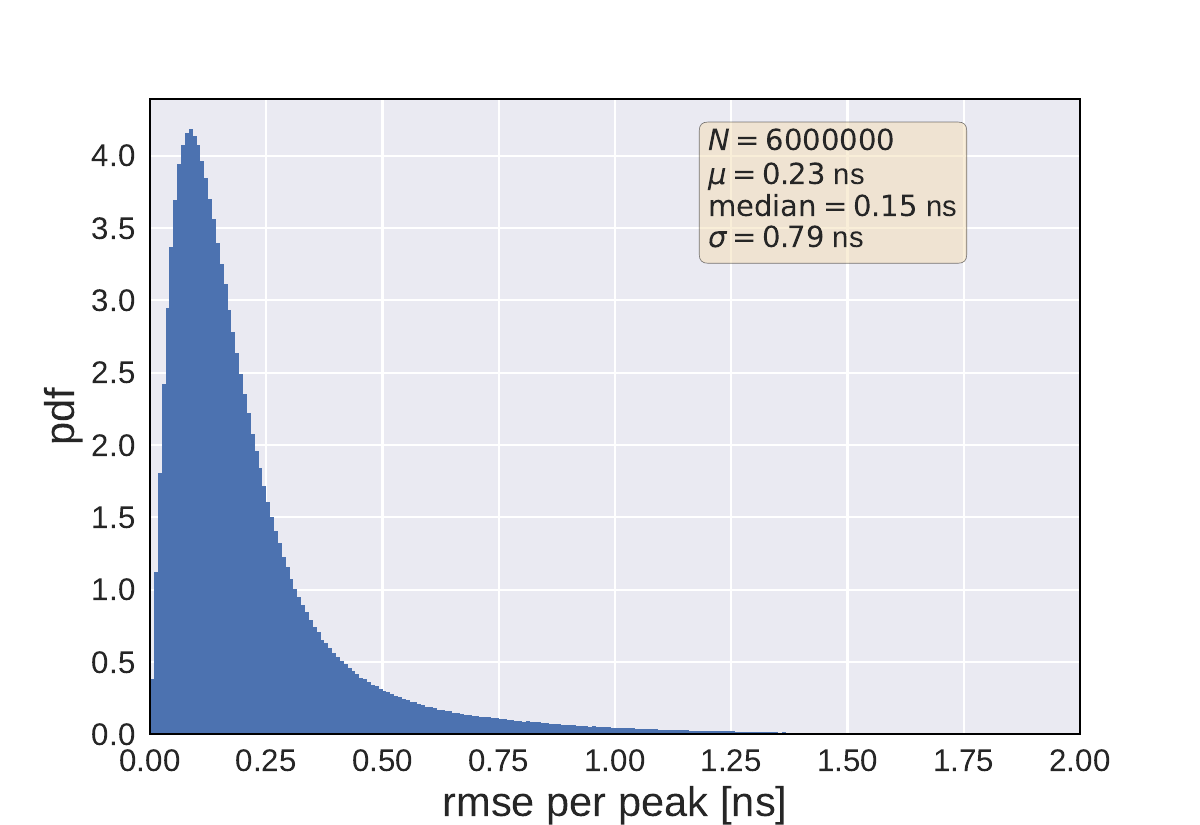}
    }
    \hfill
    \subfigure[]{
    \includegraphics[width=0.46\textwidth,  trim={2cm, 0.5cm, 2cm, 0.7cm}]{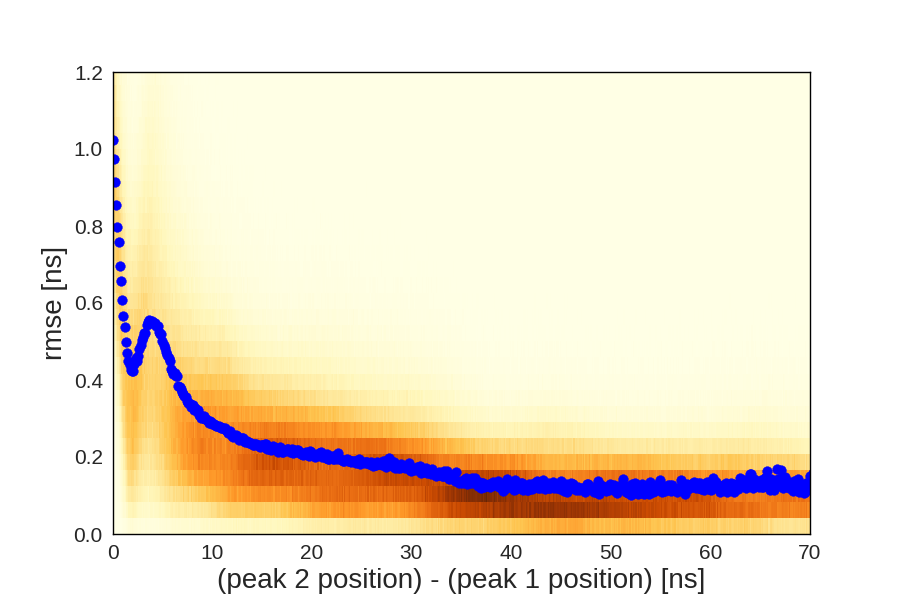}
    }
    \caption{Deep Double Peak Finder (without MCP): (a) RMSE distribution. (b) RMSE as a function of the peak distance. Performance of the peak finder model on 6M simulated double hit signals from all six channels.
    In (b) the background shows the 2D histogram normalized such that each column adds up to 1 and the blue dots show the mean at the respective peak distance. 
    }
    \label{fig:peak_finder_performance}
\end{figure}

\begin{figure}[t]
    \centering
    \includegraphics[width=\textwidth]{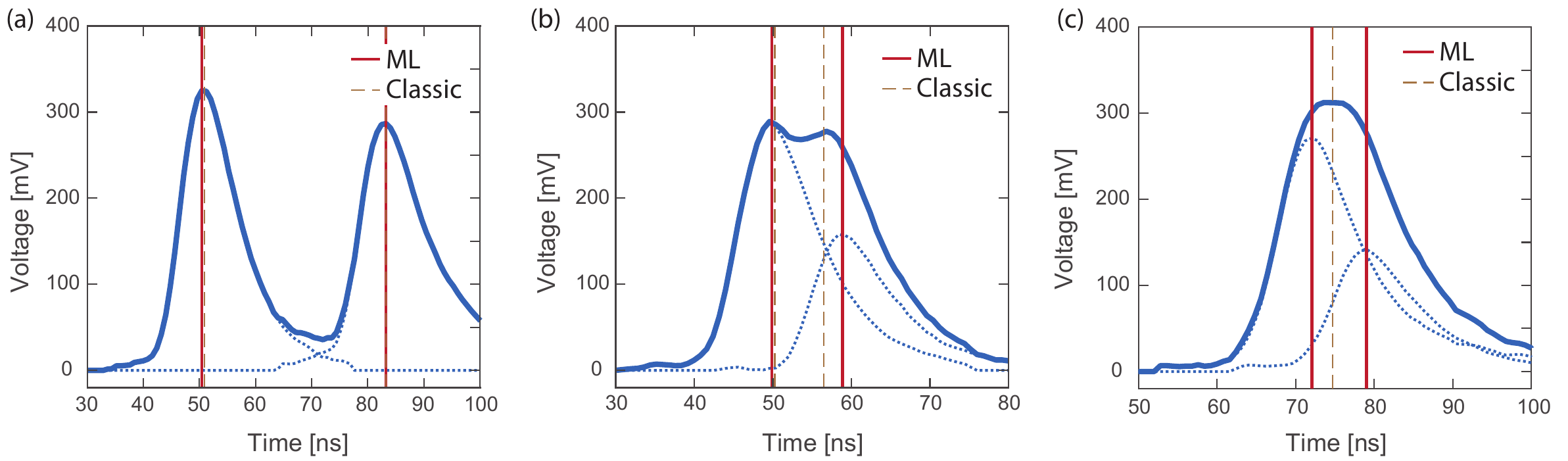}
    \caption{Example events with simulated constructed double events. Thick blue line: the data that the classical and the machine learning algorithm received. It consists of the two thin blue lines, so in this case we can exactly know where the peaks should be. (a) Event with two distinct peaks. The classical as well as the machine learning algorithm can capture both peaks well. (b) Event with two close peaks: the classical algorithm still sees two peaks, but they are reconstructed in at the wrong positions as both traces strongly overlap. (c) very close event: the machine learning algorithm is still capable to find both peak positions, while the classical algorithm only finds one maximum which is not at any of the real peak positions.}
    \label{fig:peaks_addition}
\end{figure}

Figure~\ref{fig:peak_finder_performance} shows the error distribution on 6M simulated double peak signals from all six channels (no MCP).
We observe an overall small error below 0.3\,ns for peaks separated by more than 10\,ns and an error below 0.6\,ns for closer hits.
The mean error increases for smaller peak distances as well as the variance. 
Closer peak positions are more difficult to determine, which is to be expected.
Figure~\ref{fig:peaks_addition} shows simulated double hit signals with their underlying single hit signals.
Once the peaks get close enough, the position of the two underlying peaks will usually no longer coincide with the maxima of the curve.

\subsection{Deep Double Peak Finder Results (MCP)}
\begin{figure}[t!]
    \centering
    \subfigure[]{\includegraphics[width=0.46\textwidth, trim={2cm, 0.5cm, 2cm, 0.5cm}]{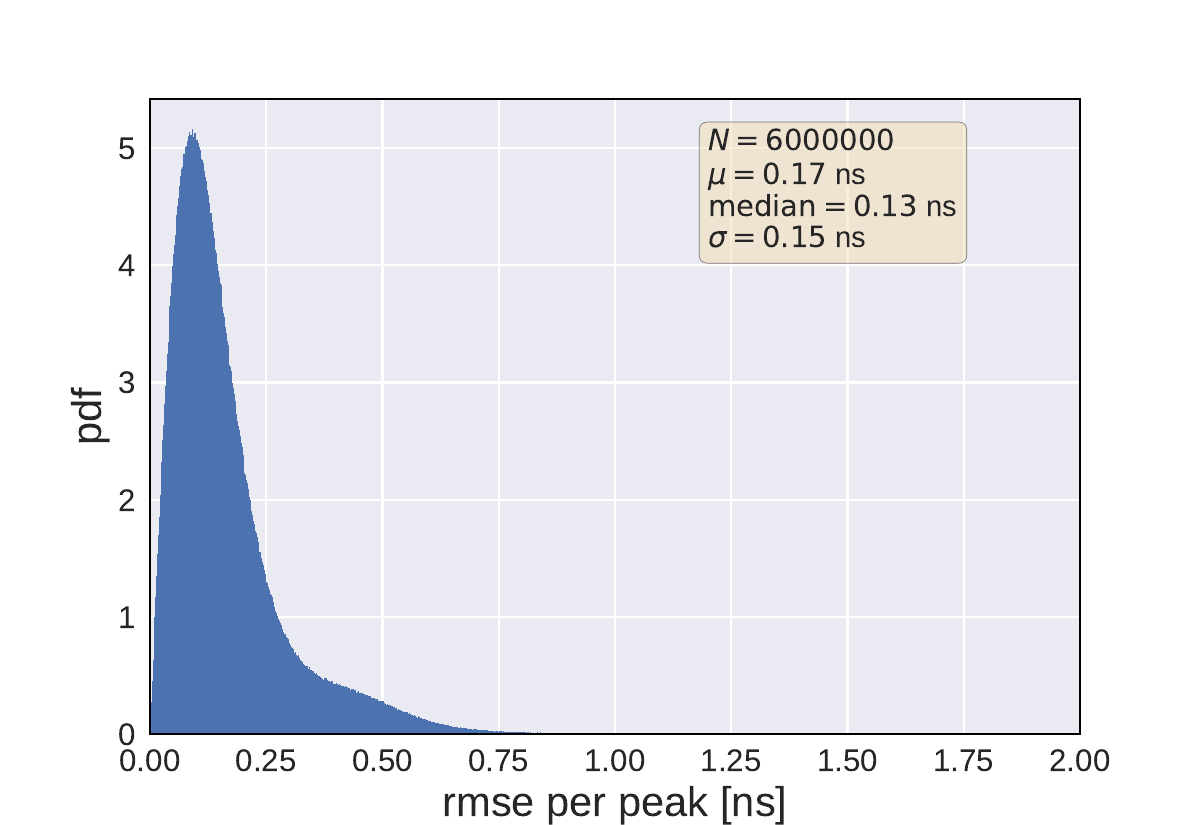}}
    \hfill
    \subfigure[]{\includegraphics[width=0.46\textwidth, trim={2cm, 0.5cm, 2cm, 0.7cm}]{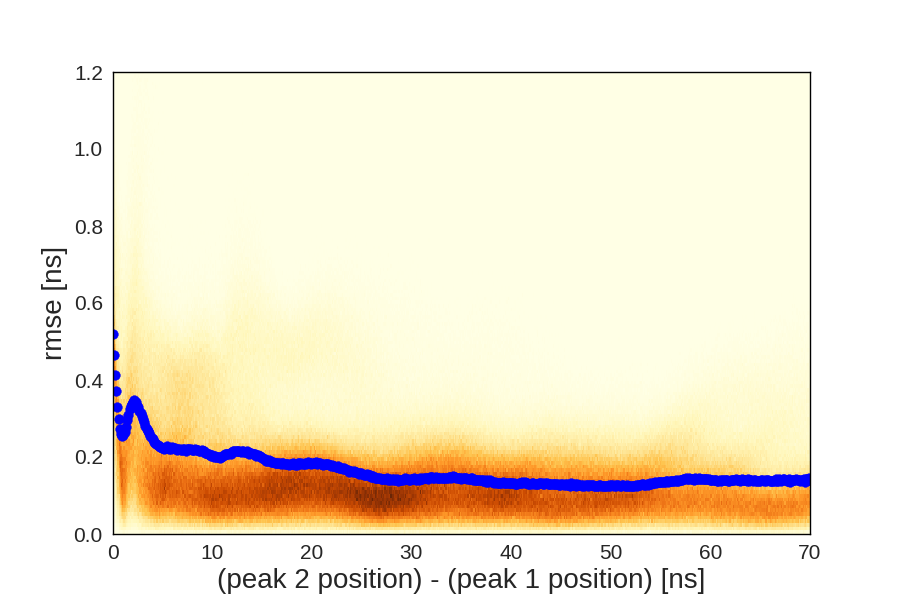}}
    \caption{Performance of the MCP only Deep Peak Finder model on simulated test data: (a) Root mean square error distribution. (b) Root mean square error as a function of the peak distance. 
    The test data consists of of 30k simulated doubles.
    In (b) the background shows the 2D histogram normalized such that each column adds up to 1 and the blue dots show the mean at the respective peak distance. 
    The mean error increases for smaller peak distances as well as the variance. Closer peaks are more difficult to accurately determine, which is to be expected.
    The outliers come from not enough data for the respective bin, 
    where one single bad prediction strongly influences the mean. 
    }
    \label{fig:mcp_model_stats}
\end{figure}

The test data consists of 6M simulated MCP double peak signals and the root mean square error is $\mathrm{rmse}_\mathrm{train} = 0.17~\mathrm{ns}$ and $\mathrm{rmse}_\mathrm{test} = 0.17~\mathrm{ns}$.
The mean average error is $\mathrm{mae}_\mathrm{train} = 0.15~\mathrm{ns}$ and $\mathrm{mae}_\mathrm{test} = 0.15~\mathrm{ns}$.
The bin size of the data is 0.8~ns.
Figure~\ref{fig:mcp_model_stats} shows the RMSE distribution (a) and the RMSE as a function of peak distance (b). The performance of the MCP model is better than the performance of the general deep double peak finder model from Figure~\ref{fig:peak_finder_performance}.

\subsection{Comparison between Classical and Machine Learning Reconstruction Models}
\begin{figure}[tb]
    \centering
    \includegraphics[width=0.75\textwidth]{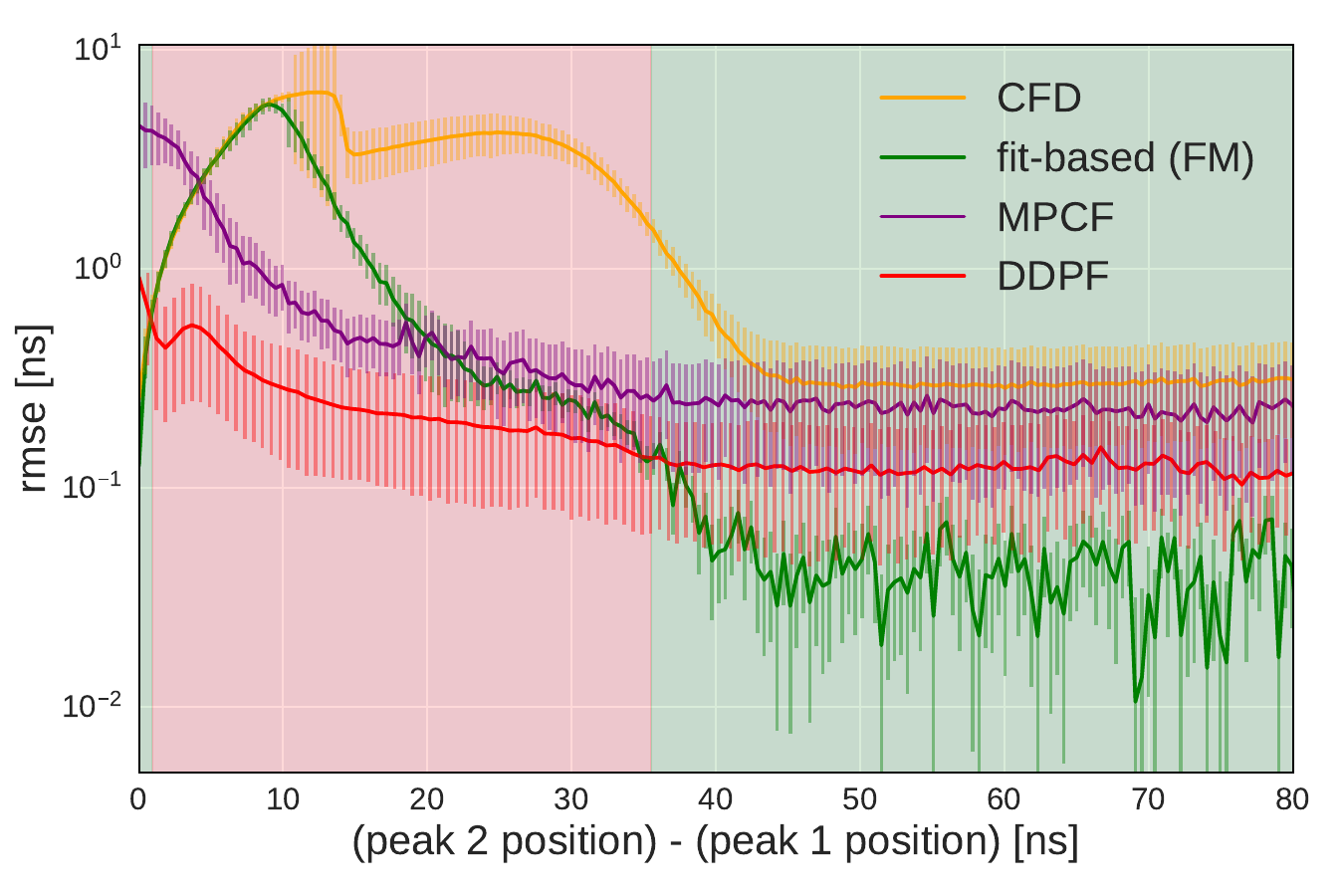}
    \caption{Log plot of the root mean squared error as a function of the peaks separation for the channels (not MCP).
    The green region is where the fit-based method has a smaller error, the red region is where the Deep Peak Finder has a smaller error.
    }
    \label{fig:ML-CL-comparison}
\end{figure}

In Figure~\ref{fig:ML-CL-comparison}, we compare the machine learning model (6 channels) to the fit-based classical algorithm described in Section~\ref{sec:fit-based-method}.
The test data consists of 3M generated signals from channels 1--6.
We filtered the data such that no peaks are closer than $16$~ns to one of the borders of the interval,
because most methods do not work when a large part of the peak is cut off.
For the error bars we fit a function to the error distribution in each peak separation bin.
The error bars correspond to one standard deviation of the fitted distribution.
We tested different functions and a Rayleigh distribution worked most consistently across all peak separations.

The classical model (shown in green) performs almost perfectly for distances $>30~\mathrm{ns}$.
This is exactly as expected, as the labels were created using the classical algorithm to find the singles and their peak positions. 
For peak distances smaller than around $22\,\mathrm{ns}$ (about the width of the peaks), the error of the classical model significantly increases, while the machine learning model performance only slightly degrades.
For much closer peaks, the classical algorithm becomes better again.
This also makes sense, as the classical algorithm only predicts one peak once the two peaks have merged.
If the peaks merge to one peak at exactly the same position, then the classical algorithm gives the correct result again (accidentally), 
without knowing that there are two peaks.
The same applies to the CFD method and the decrease in RMSE for peak separations closer than 10--20~ns for CFD and fit-based is purely this effect. If all the signals where only one peak is predicted are dropped,
then this effect goes away.

The huge difference between RMSE and MAE in the peak finding models comes from few very large errors, as RMSE puts a higher weight on large errors.
If we take out all data points where the error is larger than $4~\mathrm{ns}$, which amounts to 0.4\% of the data, then $\mathrm{rmse}_\mathrm{test} = 0.30~\mathrm{ns}$ and $\mathrm{mae}_\mathrm{test} = 0.06~\mathrm{ns}$.
If the position calculation software cannot reconstruct the event, it is typically discarded.
The higher test loss than training loss indicates some amount of overfitting.
Increasing dropout did not increase model performance on validation data.

\begin{figure}[tb]
   \includegraphics[width=\textwidth, clip, trim=3.5cm 0cm 0.5cm 0cm]{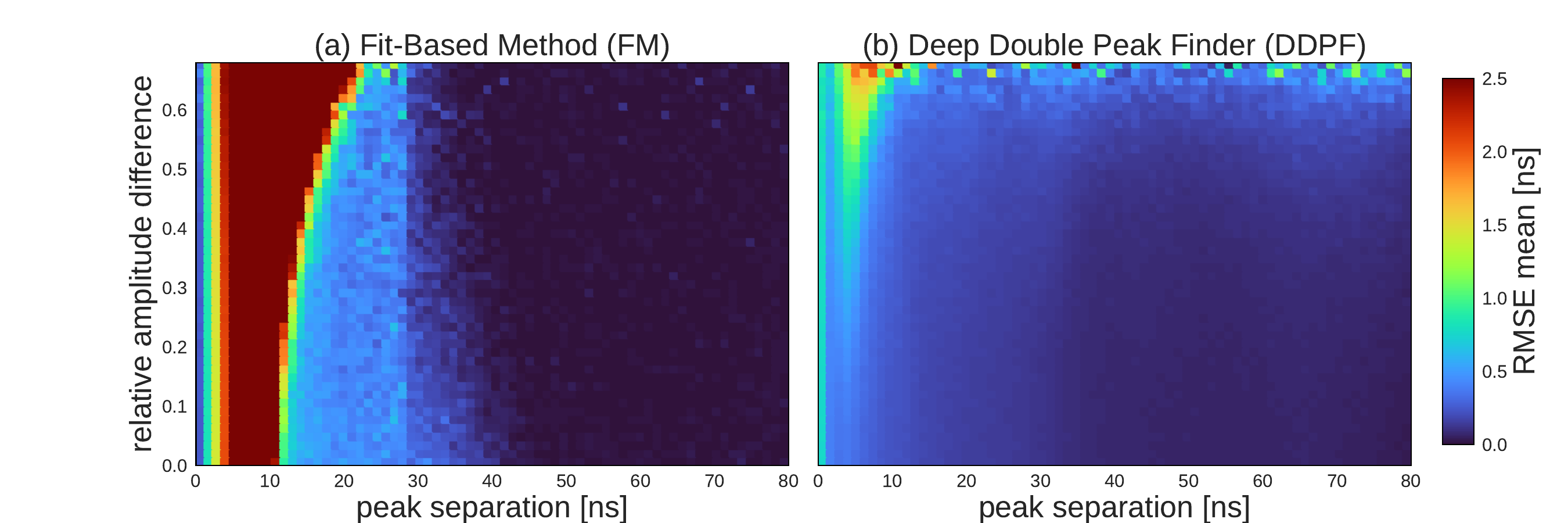} 
    \caption{RMSE mean \new{(color)} as a function of peak separation and relative amplitudes difference. \new{(a) Fit-Based Method, (b) Deep Double Peak Finder.}}
    \label{fig:rel_ampl_diff_hist}
\end{figure}

\begin{figure}[tb]
    \centering
    \includegraphics[width=0.7\textwidth]{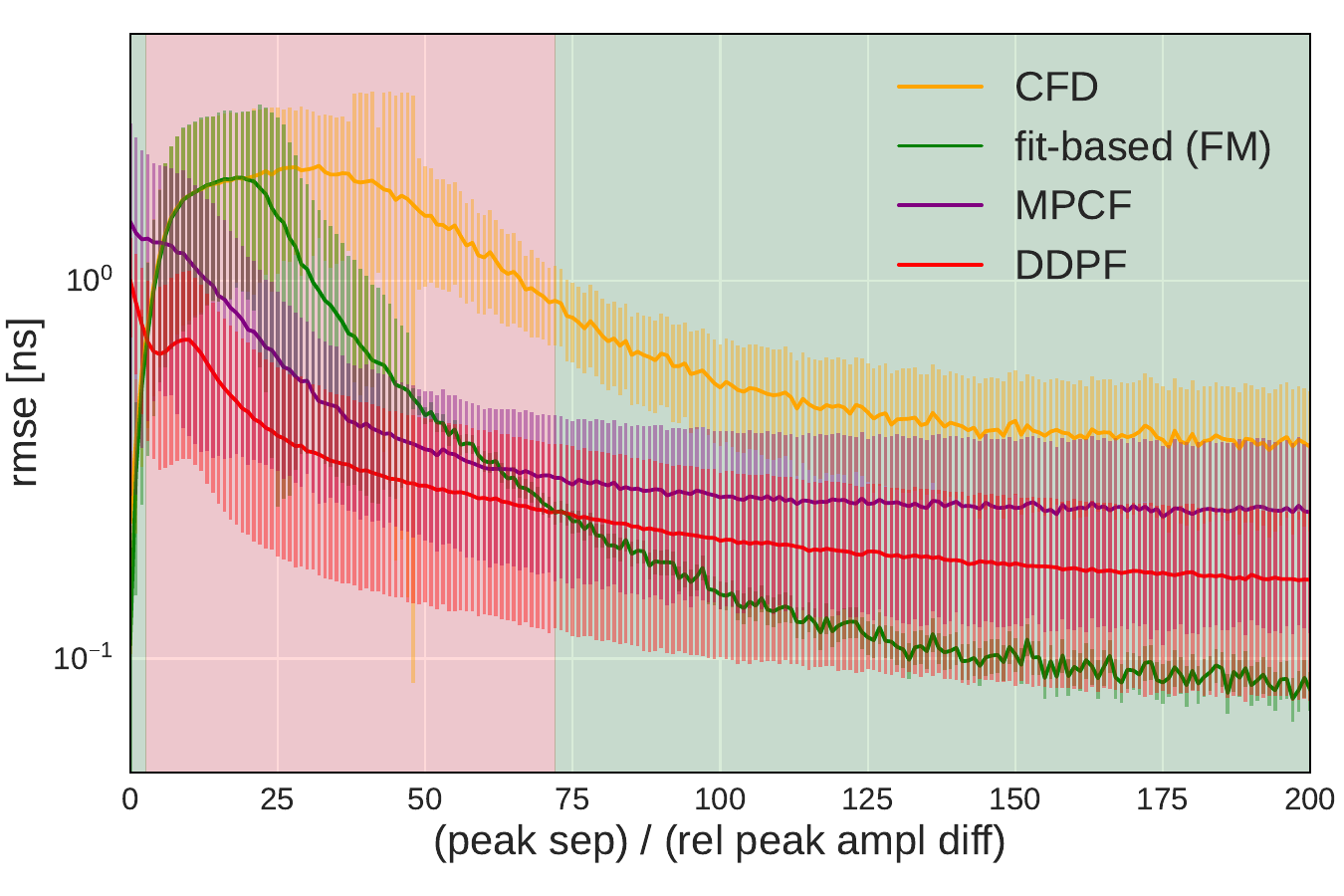}
    \caption{RMSE comparison between the different methods as a function of
    $\xi = \frac{\mathrm{peak\ separation}}{\mathrm{rel.\ ampl.\ diff.}}$}.
    The green region is where the fit-based method has a smaller error, the red region is where the Deep Peak Finder has a smaller error.
    \label{fig:rmse_comparison_xi}
\end{figure}

The performance of the models not only depends on the peak separation, but also on the amplitudes $a_0$ and $a_1$ of the underlying signals.
Two peaks of about the same height can get closer to each other without merging to a single peak than two peaks with very different heights can.
In order to investigate the performance of the models with respect to the differences in amplitudes, we introduce the \say{relative amplitudes difference} metric: $|a_0 - a_1|/(a_0+a_1)$.
Figure~\ref{fig:rel_ampl_diff_hist} shows a \old{3D} plot where the peak separation and the relative amplitudes difference are on the $x$- and $y$-axis and the \old{$z$-axis contains} \new{color shows} the mean of the RMSE distribution for the respective bin.
We can now introduce a function of the peak separation and the relative amplitudes difference and reduce the plots in Figure~\ref{fig:rel_ampl_diff_hist} to two dimensions.
We introduce
\begin{equation}
    \xi = \frac{\mathrm{peak\ separation}}{\mathrm{rel.\ ampl.\ diff.}}\,,
\end{equation}
motivated by the premise that it is hard to predict the peak positions when
a) the peaks have a small separation and b) when the relative amplitudes difference is large. 
Thus the region with small $\xi$ is hard to predict and the region with large $\xi$ is much easier to predict, as shown in Figure~\ref{fig:rmse_comparison_xi}.
Constant $\xi$ corresponds to a straight line through the origin in the
$(\mathrm{rel.\ ampl.\ diff.})$-$(\mathrm{peak\ separation})$-plane and a point in
Figure~\ref{fig:rmse_comparison_xi} corresponds to the mean along the respective line.
The error bars are from fitting a Rayleigh distribution to the RMSE distribution in the respective $\xi$-bin and taking one standard deviation of the fitted distribution.



\subsection{Real Data Inference}
\label{sec:stateval}

\begin{figure}[ht!]
    \centering
    \includegraphics[width=\textwidth]{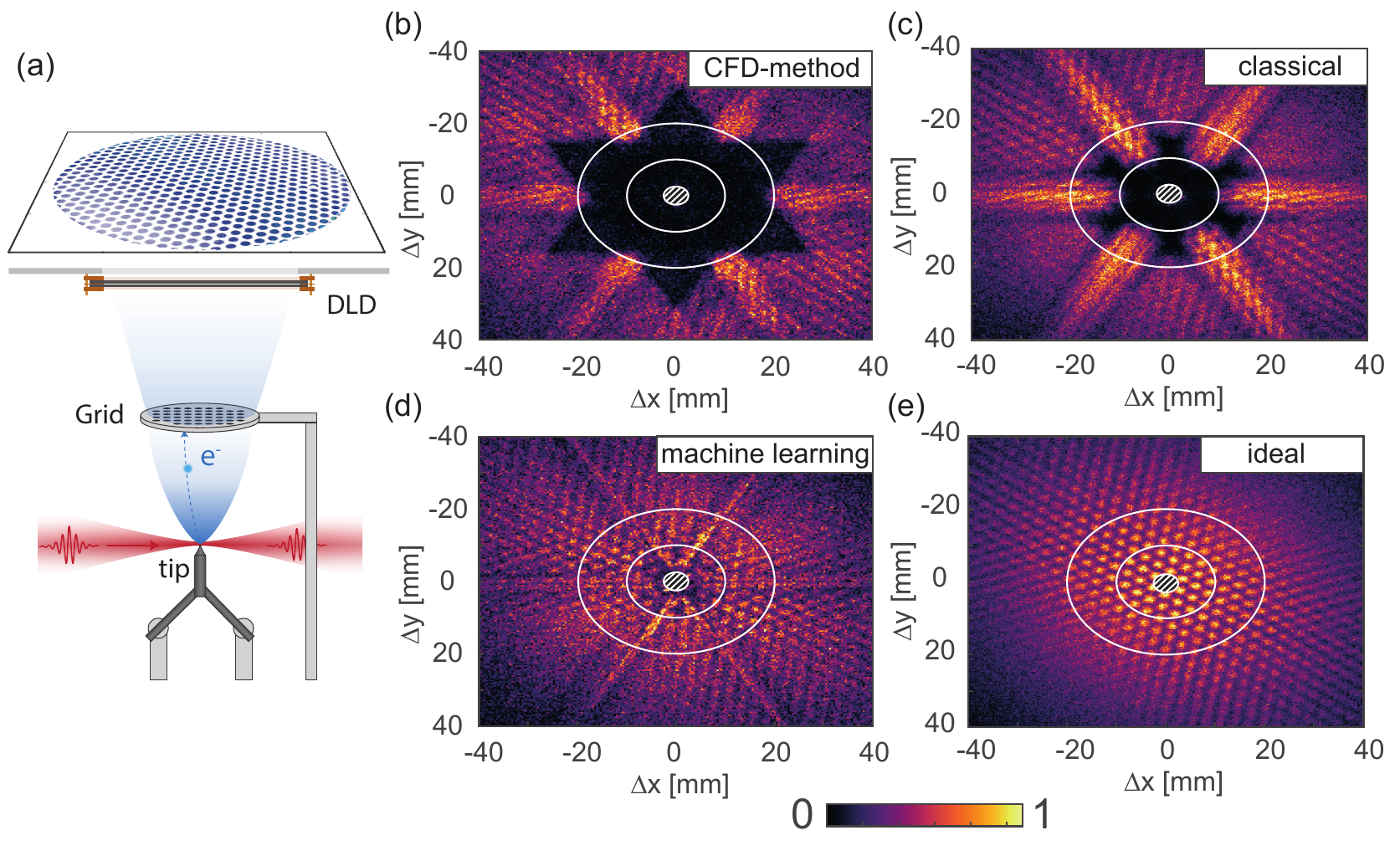}
    \caption{Position difference plots for data with a grid in front of the detector, casting a clear and regular electron shadow on the detector. $\Delta x$ is $x_2-x_1$, where $x_i$ is the position of particle $i$, $\rmd y$ respectively.
    (a) Schematic display of the experimental setup. 
    (b) positions from Constant Fraction Discriminator method,
    (c) positions from fit-based classical algorithm,
    (d) positions from the machine learning model with the Deep Double Peak Finder model,
    (e) \say{ideal} plot purely from two uncorrelated singles.
    The two white circles indicate 10\,mm and 20\,mm on the images for comparison.
    The inner white circle with the white dashed area corresponds to the dead-area of the DDPF.
    }
    \label{fig:grids}
\end{figure}

As a final check, we compare the classical evaluation methods that are given by the hardware-based CFD peak retrieval method and the fit-based classical method to the results we obtained with the machine learning models described in Section \ref{sec:ML_setup} using real data.
\new{
The following evaluation and the given dead-radii are for true simultaneous double hits, meaning most electron pairs arrive within one ns.
If the arrival time of the particles is widely spread out, also the dead-radius will get smaller as the particles can get distinguished by the time of flight and their signals will generally overlap less.}
\old{To do that} \new{W}e perform a measurement in which a copper grid, commercially available for sample preparation in transmission electron microscopy, is placed in front of the electron source as shown in Figure~\ref{fig:grids}.
The shadow image of the grid is projected on the DLD, as shown in Figure~\ref{fig:grids} (a).
Thus, there are sharp spatial features, either in real space for the first and the second electron, but also for the difference plots in which \(\Delta y\) is plotted over \(\Delta x\) for the double hits.
The clarity of the grid appearing in the reconstructed positions and difference plots is an indicator for the quality of the method's performance.
In the center of these images, a white circle with white lines is plotted that shows the dead region of the DDPF that we cut out in all four cases.
In this region, it is not possible anymore to distinguish between two hits of low amplitude or one hit of large amplitude, which is why there errors will occur.

Figure~\ref{fig:grids} (b,c,d,e) shows the position differences, i.e.\ 
a 2D histogram of the $x$ and $y$ distances of the two particles.
Figure~\ref{fig:grids}~(b) shows double hit events evaluated with the CFD-method.
There is a star-shaped dead region (black region) with a maximal extension of about $30\,\mathrm{mm}$ where most events cannot be reconstructed.
There are also other weaker 6-fold artifacts.
Figure~\ref{fig:grids} (c) shows double hit events evaluated with the fit-based classical peak finding algorithm.
There are strong 6-fold artifacts, too, and the dead radius is at about $20\,\mathrm{mm}$.
The strong hexagonal artifacts originate from the increasingly poorly determined peak positions as soon as the two signals come very close to each other, as shown in Figure~\ref{fig:peaks_addition}.
The associated signals of the individual electrons on the different layers cannot be clearly assigned by the evaluation software, which subsequently leads to these artifacts.
Figure~\ref{fig:grids}~(d) shows the same double events evaluated with the Deep Double Peak Finder.
While some remaining 6-fold artifacts are still visible, they are much weaker now compared to the classical methods in b) and c).
d decrease the dead radius to a much smaller value of \(\sim2.5\,\)mm and the grid appears much sharper in the region closer to the center.

Figure~\ref{fig:grids} (e) shows the difference between two single hit events. This is an "ideal" plot obtained from two uncorrelated single hit events, showing a very good resolution of the grid. 
There, no intrinsic dead radius or other double-hit based detector artifacts as well as possible physical correlations are present.


The absolute numbers for the dead radius entirely depend on the FWHM of the pulse, here 12.5\,ns (evaluated for a typical norm pulse). If the width of the norm pulse can be reduced by half through hardware changes, for example, then all dead radii can also be reduced by half, both those of the hardware-based evaluation and those of the classical algorithm as well as the machine-learning-based evaluation.
This can be seen in the original article to the multihit capability of the DLD, see Ref.~\cite{Jagutzki2002_2}, where the authors obtained a FWHM of the pulses of 4-5\,ns, thus already obtaining a dead radius of \(<10\,\)mm.
Since our evaluation enables a relative improvement compared to the peak width, such a signal width would also further improve our dead radius.
Typically, also the particle species can influence the peak width, e.g.\ it should be smaller for ions. 
\old{
All given dead-radii are for true simultaneous hits, so two electrons that arrive within a few ns or even below (even \(<1\)\,ns in the case of the data shown in Fig.~\ref{fig:grids}). If the arrival time of the particles is widely spread out, also the dead-radius will get smaller as the particles can get distinguished by the time of flight.
}


\section{Interpretation and Discussion}
\label{sec:interpretation}
As expected, the worst results are obtained using the hardware-based peak detection method with the CFD method.
There is only one detectable zero crossing when two peaks come too close to each other.
Therefore, we expect a large dead radius, which we observed in the difference plot in Figure~\ref{fig:grids}~(b).
Additionally, the grid is barely visible in this plot.
By comparison, in Figure~\ref{fig:grids}~(e), we see how the difference plot would look like for perfectly resolved double hits:
the grid in real space results in a nicely observable pattern in the difference plot.

For peaks that are far enough apart, the fit-based classical algorithm is perfect and beats the DDPF, since the peaks do not influence each other and the labels were created with help of the classical algorithm as shown in Figure~\ref{fig:ML-CL-comparison}.
However, the difference between the machine learning model and the fit-based classical algorithm for well separated peaks ($>20$ ns) is on the order of 1/4 of the bin size (see Figure~\ref{fig:peak_finder_performance}(a) and Figure~\ref{fig:mcp_model_stats}(a)) and so the machine learning model still performs reasonably well in this regime.
Once the peaks get closer 
they can influence the peak positions of the added signal.
The peaks typically have a mean width of 12.5\,ns (FWHM), so when the temporal separation comes close to this width, we expect the peak finding to become harder for the classical methods as well as for the machine learning model, as shown in Figure~\ref{fig:peaks_addition}~(b).
Already starting from 25--30\,ns separation and below, we see an increase in RMSE in Figure~\ref{fig:ML-CL-comparison} for the classical method
and the machine learning model performs visibly better below 25\,ns.
If the peaks get close, it often happens that there is only one single maximum in the added signal (Figure~\ref{fig:peaks_addition}~(c)).
The merging into one single maximum depends on the distance between the peaks and also the two amplitudes, so no hard threshold can be given. 
Below $\approx$ 10\,ns peak separation the RMSE of the classical method strongly increases in Figure~\ref{fig:ML-CL-comparison}.
We also see it in the right panel in Figure~\ref{fig:peaks_addition}.
The classical algorithm only finds one peak, whereas the machine learning model finds two peaks that fit well to the underlying peaks. 
Only for the case where the two peaks almost are on top of each other, below a distance of 1\,ns in Figure~\ref{fig:ML-CL-comparison},
the classical algorithm performs better again.
This is expected since the classical algorithm finds the one peak and it is treated as two at the same position,
while the ML model tries to find two peaks that would fit.
Note that here the classical algorithm succeeds accidentally, as it believes there is only a single peak. Even in this region, the machine learning model performs well and the error is close to or below the bin size. Overall, the machine learning model performs better that the classical methods for peaks that are closer than 25--30\,ns separation down to 1\,ns. 
How easily the peak positions can be determined not only depends on the peak separation, but also on the relative amplitudes of the two underlying peaks as illustrated in Figures~\ref{fig:rel_ampl_diff_hist} (a) and (b).
The fit-based method performs better for far peak separations for all values of the relative amplitude difference, at closer peak separations it gets very inconsistent, whereas the DDPF has a much smoother RMSE-mean-surface and stays lower and more consistent for closer peak separations.
Combining peak separation and relative amplitude difference to a single variable $\xi$, Figure~\ref{fig:rmse_comparison_xi} also shows that for $2.7<\xi<72$~[ns] the DDPF outperforms 
the other methods.
Again, for $\xi<2.7$~[ns] the fit-based method's rmse spuriously drops drastically, since here it only predicts one single peak and we had to take it as a double peak at this position in order to do this evaluation.

Furthermore, based on the grid plots obtained with real data in Figure~\ref{fig:grids}, it is clear that the Deep Double Peak Finder model performs better than the classical methods on real data, showing a much smaller dead radius, less artifacts and better resolution compared to classical methods. While this was to be expected from the comparison between real and simulated doubles in Figure~\ref{fig:artificial_real_double_comparison}, the grid plots further confirms this conclusion.

Note that in our exploration of various architectures, the Convolutional Neural Network (CNN) demonstrated notable efficacy as a classifier.
However, its performance in peak detection tasks was less than satisfactory.
Models based on Gated Recurrent Units (GRUs) exhibited superior performance in identifying peak positions.
The use of Long Short-Term Memory (LSTM) networks, instead of GRUs, did not yield any discernible improvement in our tests.


\section{Summary and Outlook}
\label{sec:summary}

This paper presents a proof of concept for improving multi-hit capabilities of delay line detectors using machine learning.
To our knowledge, it is the first time that the spatial and temporal resolution of a delay line detector has been improved with the application of machine learning algorithms.

We have shown that a machine learning approach that consists of Hit Multiplicity Classifier and Deep Double Peak Finder models performs well on simulated and real data.
The machine learning model finds the correct peak multiplicities and peak positions to a good accuracy even when classical methods fail to identify multi-particle hits. The models were also evaluated on real data using the grid approach as a point-projection image with an ultrafast low-energy electron source. This measurement, summarized in Figure~\ref{fig:grids}, can serve as a figure of merit for the quality of two-electron event evaluation on real data.
It shows that our machine learning model surpasses the classical algorithms notably by having a smaller dead radius, fewer artifacts and a better resolution of the grid.

The current machine learning model is trained to identify multilple hit signals and only reconstruct double hit signals.
The Hit Multiplicity Classifier already works up to quadruple hits with AUC $>0.9998$. In future follow-up work, we plan to extend the Deep Peak Finder to triple and quadruple events. We expect that our general method---generating multi-peak data through addition of singles, using a classifier followed by a peak finder for the specific multiplicity---will work for all or most applications where the signals are additive or approximately additive.

As this evaluation is applicable for any existing delay line-based detector with the possibility of analog read-out, it is highly relevant for many ultrafast pulsed applications, including experiments on  nonsequential double ionization~\cite{Weber2000,Becker2012}, atom probe tomography~\cite{Costa2012}, ultrafast electron microscopy~\cite{Hassan2017,Shiloh2022} or general correlation experiments like of the fermionic Hanbury Brown-Twiss experiment~\cite{Kiesel2002,Kuwahara2021}.
Further plans include prediction of the particle positions and times instead of only predicting the peak positions,
implementation of machine learning models in the online regime
and other model architectures.
Currently, some first experimental works are investigating towards the usage of triplet and quadruplet events \cite{Basnayake2021,Meier2023,Haindl2023}.
Therefore, higher hit event multiplicity reconstruction will be a promising route for future work.

\section{Acknowledgements}
The FAU members acknowledge support by the European Research Council (Consolidator Grant NearFieldAtto and Advanced Grant AccelOnChip) and the Deutsche Forschungsgemeinschaft (DFG, German Research Foundation)---Project-ID 429529648---TRR 306 QuCoLiMa (\say{Quantum Cooperativity of Light and Matter}) and Sonderforschungsbereich 953 (\say{Synthetic Carbon Allotropes}), Project-ID 182849149. J.H.\ acknowledges funding from the Max Planck School of Photonics.
This work was supported in part by the U.S.\ Department of Energy (DOE) under Award No.\ DE-SC0012447 (S.G.).
M.K.\ acknowledges support through the Graduate Council Fellowship of the University of Alabama and, in part, through the U.S.\ Department of Energy grant DE-SC0012447. 

\section*{Data Availablity Statement}
The data that support the findings of this study are available upon reasonable request from the authors.

\printbibliography

@ARTICLE{2014GRU,
       author = {{Cho}, Kyunghyun and {van Merrienboer}, Bart and {Bahdanau}, Dzmitry and {Bengio}, Yoshua},
        title = "{On the Properties of Neural Machine Translation: Encoder-Decoder Approaches}",
      journal = {arXiv e-prints},
         year = 2014,
        month = sep,
          eid = {arXiv:1409.1259},
        pages = {arXiv:1409.1259},
archivePrefix = {arXiv},
       eprint = {1409.1259},
 primaryClass = {cs.CL},
       adsurl = {https://ui.adsabs.harvard.edu/abs/2014arXiv1409.1259C}
}

@ARTICLE{2016Hyperband,
       author = {{Li}, Lisha and {Jamieson}, Kevin and {DeSalvo}, Giulia and {Rostamizadeh}, Afshin and {Talwalkar}, Ameet},
        title = "{Hyperband: A Novel Bandit-Based Approach to Hyperparameter Optimization}",
      journal = {arXiv e-prints},
         year = 2016,
        month = mar,
          eid = {arXiv:1603.06560},
        pages = {arXiv:1603.06560},
archivePrefix = {arXiv},
       eprint = {1603.06560},
 primaryClass = {cs.LG},
       adsurl = {https://ui.adsabs.harvard.edu/abs/2016arXiv160306560L}
}

@Article{Jagutzki2002,
  author =    {O. Jagutzki and V. Mergel and K. Ullmann-Pfleger and L. Spielberger and U. Spillmann and R. D\"{o}rner and H. Schmidt-B\"{o}cking},
  title =     {A broad-application microchannel-plate detector system for advanced particle or photon detection tasks: large area imaging, precise multi-hit timing information and high detection rate},
  journal =   {Nuclear Instruments and Methods in Physics Research Section A: Accelerators, Spectrometers, Detectors and Associated Equipment},
  year =      {2002},
  volume =    {477},
  number =    {1-3},
  pages =     {244--249},
  month =     jan,
  doi =       {10.1016/s0168-9002(01)01839-3},
  publisher = {Elsevier {BV}},
  url =       {https://doi.org/10.1016/s0168-9002(01)01839-3}
}

@Article{Ullrich2003,
  author =    {J. Ullrich and R. Moshammer and A. Dorn and R. D\"{o}rner and L. Ph. H .Schmidt and H. Schmidt-B\"{o}cking},
  title =     {Recoil-ion and electron momentum spectroscopy: reaction-microscopes},
  journal =   {Reports on Progress in Physics},
  year =      {2003},
  volume =    {66},
  number =    {9},
  pages =     {1463--1545},
  month =     aug,
  doi =       {10.1088/0034-4885/66/9/203},
  publisher = {{IOP} Publishing},
  url =       {https://doi.org/10.1088/0034-4885/66/9/203}
}

@Article{Weber2000,
  author =    {Th. Weber and H. Giessen and M. Weckenbrock and G. Urbasch and A. Staudte and L. Spielberger and O. Jagutzki and V. Mergel and M. Vollmer and R. D\"{o}rner},
  title =     {Correlated electron emission in multiphoton double ionization},
  journal =   {Nature},
  year =      {2000},
  volume =    {405},
  number =    {6787},
  pages =     {658--661},
  month =     jun,
  doi =       {10.1038/35015033},
  publisher = {Springer Science and Business Media {LLC}},
  url =       {https://doi.org/10.1038/35015033}
}

@Article{Jeltes2007,
  author =    {T. Jeltes and J. M. McNamara and W. Hogervorst and W. Vassen and V. Krachmalnicoff and M. Schellekens and A. Perrin and H. Chang and D. Boiron and A. Aspect and C. I. Westbrook},
  title =     {Comparison of the Hanbury Brown{\textendash}Twiss effect for bosons and fermions},
  journal =   {Nature},
  year =      {2007},
  volume =    {445},
  number =    {7126},
  pages =     {402--405},
  month =     jan,
  doi =       {10.1038/nature05513},
  publisher = {Springer Nature},
  url =       {https://doi.org/10.1038/nature05513}
}

@Article{Kuwahara2021,
  author =    {Makoto Kuwahara and Yuya Yoshida and Wataru Nagata and Kojiro Nakakura and Masato Furui and Takafumi Ishida and Koh Saitoh and Toru Ujihara and Nobuo Tanaka},
  title =     {Intensity Interference in a Coherent Spin-Polarized Electron Beam},
  journal =   {Physical Review Letters},
  year =      {2021},
  volume =    {126},
  number =    {12},
  pages =     {125501},
  month =     mar,
  doi =       {10.1103/physrevlett.126.125501},
  publisher = {American Physical Society ({APS})},
  url =       {https://doi.org/10.1103/physrevlett.126.125501}
}

@article{Gys2015,
  doi = {10.1016/j.nima.2014.12.044},
  url = {https://doi.org/10.1016/j.nima.2014.12.044},
  year = {2015},
  month = jul,
  publisher = {Elsevier {BV}},
  volume = {787},
  pages = {254--260},
  author = {T. Gys},
  title = {Micro-channel plates and vacuum detectors},
  journal = {Nuclear Instruments and Methods in Physics Research Section A: Accelerators,  Spectrometers,  Detectors and Associated Equipment}
}

@Article{Lubsandorzhiev2006,
  author =    {B.K. Lubsandorzhiev},
  title =     {On the history of photomultiplier tube invention},
  journal =   {Nuclear Instruments and Methods in Physics Research Section A: Accelerators, Spectrometers, Detectors and Associated Equipment},
  year =      {2006},
  volume =    {567},
  number =    {1},
  pages =     {236--238},
  month =     nov,
  doi =       {10.1016/j.nima.2006.05.221},
  publisher = {Elsevier {BV}},
  url =       {https://doi.org/10.1016/j.nima.2006.05.221}
}

@Article{Kiesel2002,
  author =    {Harald Kiesel and Andreas Renz and Franz Hasselbach},
  title =     {Observation of Hanbury Brown{\textendash}Twiss anticorrelations for free electrons},
  journal =   {Nature},
  year =      {2002},
  volume =    {418},
  number =    {6896},
  pages =     {392--394},
  month =     jul,
  doi =       {10.1038/nature00911},
  publisher = {Nature Publishing Group},
  url =       {http://dx.doi.org/10.1038/nature00911}
}

@article{CMS:2022wjj,
    collaboration = "CMS",
    title = "{Reconstruction of decays to merged photons using end-to-end deep learning with domain continuation in the CMS detector}",
    eprint = "2204.12313",
    archivePrefix = "arXiv",
    primaryClass = "hep-ex",
    reportNumber = "CMS-EGM-20-001, CERN-EP-2022-028",
    month = "4",
    year = "2022"
}

@ARTICLE{2020SciPy-NMeth,
  author  = {Virtanen, Pauli and Gommers, Ralf and Oliphant, Travis E. and
            Haberland, Matt and Reddy, Tyler and Cournapeau, David and
            Burovski, Evgeni and Peterson, Pearu and Weckesser, Warren and
            Bright, Jonathan and {van der Walt}, St{\'e}fan J. and
            Brett, Matthew and Wilson, Joshua and Millman, K. Jarrod and
            Mayorov, Nikolay and Nelson, Andrew R. J. and Jones, Eric and
            Kern, Robert and Larson, Eric and Carey, C J and
            Polat, {\.I}lhan and Feng, Yu and Moore, Eric W. and
            {VanderPlas}, Jake and Laxalde, Denis and Perktold, Josef and
            Cimrman, Robert and Henriksen, Ian and Quintero, E. A. and
            Harris, Charles R. and Archibald, Anne M. and
            Ribeiro, Ant{\^o}nio H. and Pedregosa, Fabian and
            {van Mulbregt}, Paul and {SciPy 1.0 Contributors}},
  title   = {{{SciPy} 1.0: Fundamental Algorithms for Scientific
            Computing in Python}},
  journal = {Nature Methods},
  year    = {2020},
  volume  = {17},
  pages   = {261--272},
  adsurl  = {https://rdcu.be/b08Wh},
  doi     = {10.1038/s41592-019-0686-2},
}

@article{LSTM,
    author = {Hochreiter, Sepp and Schmidhuber, Jürgen},
    title = "{Long Short-Term Memory}",
    journal = {Neural Computation},
    volume = {9},
    number = {8},
    pages = {1735-1780},
    year = {1997},
    month = {11},
    issn = {0899-7667},
    doi = {10.1162/neco.1997.9.8.1735},
    url = {https://doi.org/10.1162/neco.1997.9.8.1735},
    eprint = {https://direct.mit.edu/neco/article-pdf/9/8/1735/813796/neco.1997.9.8.1735.pdf},
}

@misc{https://doi.org/10.48550/arxiv.2011.10616,
  doi = {10.48550/ARXIV.2011.10616},
  url = {https://arxiv.org/abs/2011.10616},
  author = {Wang, Rui and Maddix, Danielle and Faloutsos, Christos and Wang, Yuyang and Yu, Rose},
  title = {Bridging Physics-based and Data-driven modeling for Learning Dynamical Systems},
  publisher = {arXiv},
  year = {2020},
  copyright = {Creative Commons Attribution 4.0 International}
}

@inproceedings{Shi_2019,
	doi = {10.1109/icra.2019.8794351},
	url = {https://doi.org/10.1109\%2Ficra.2019.8794351},
	year = 2019,
	month = may,
	publisher = {{IEEE}},
	author = {Guanya Shi and Xichen Shi and Michael O{\textquotesingle}Connell and Rose Yu and Kamyar Azizzadenesheli and Animashree Anandkumar and Yisong Yue and Soon-Jo Chung},
	title = {Neural Lander: Stable Drone Landing Control Using Learned Dynamics},
	booktitle = {2019 International Conference on Robotics and Automation ({ICRA})}
}

@misc{DeepGLEAM,
  doi = {10.48550/ARXIV.2102.06684},
  url = {https://arxiv.org/abs/2102.06684},
  author = {Wu, Dongxia and Gao, Liyao and Xiong, Xinyue and Chinazzi, Matteo and Vespignani, Alessandro and Ma, Yi-An and Yu, Rose},
  title = {DeepGLEAM: A hybrid mechanistic and deep learning model for COVID-19 forecasting},
  publisher = {arXiv},
  year = {2021},
  copyright = {arXiv.org perpetual, non-exclusive license}
}

@misc{omalley2019kerastuner,
	title        = {Keras {Tuner}},
	author       = {O'Malley, Tom and Bursztein, Elie and Long, James and Chollet, Fran\c{c}ois and Jin, Haifeng and Invernizzi, Luca and others},
	year         = 2019,
	howpublished = {\url{https://github.com/keras-team/keras-tuner}}
}

@InProceedings{NeuralPointProcess,
  title = 	 {Neural Point Process for Learning Spatiotemporal Event Dynamics},
  author =       {Zhou, Zihao and Yang, Xingyi and Rossi, Ryan and Zhao, Handong and Yu, Rose},
  booktitle = 	 {Proceedings of The 4th Annual Learning for Dynamics and Control Conference},
  pages = 	 {777--789},
  year = 	 {2022},
  editor = 	 {Firoozi, Roya and Mehr, Negar and Yel, Esen and Antonova, Rika and Bohg, Jeannette and Schwager, Mac and Kochenderfer, Mykel},
  volume = 	 {168},
  series = 	 {Proceedings of Machine Learning Research},
  month = 	 Jun,
  publisher =    {PMLR},
  url = 	 {https://proceedings.mlr.press/v168/zhou22a.html}
}

@Article{CovarianceMatrixHierarchicalBayesianSpatioTemporal,
AUTHOR = {Sun, Bin and Wu, Yuehua},
TITLE = {Estimation of the Covariance Matrix in Hierarchical Bayesian Spatio-Temporal Modeling via Dimension Expansion},
JOURNAL = {Entropy},
VOLUME = {24},
YEAR = {2022},
NUMBER = {4},
ARTICLE-NUMBER = {492},
URL = {https://www.mdpi.com/1099-4300/24/4/492},
PubMedID = {35455155},
ISSN = {1099-4300},
DOI = {10.3390/e24040492}
}

@Article{SpatialModelingPrecipitation,
AUTHOR = {Agou, Vasiliki D. and Pavlides, Andrew and Hristopulos, Dionissios T.},
TITLE = {Spatial Modeling of Precipitation Based on Data-Driven Warping of Gaussian Processes},
JOURNAL = {Entropy},
VOLUME = {24},
YEAR = {2022},
NUMBER = {3},
ARTICLE-NUMBER = {321},
URL = {https://www.mdpi.com/1099-4300/24/3/321},
ISSN = {1099-4300},
DOI = {10.3390/e24030321}
}

@ARTICLE{PredictingClusteredWeatherPatterns,
       author = {{Chattopadhyay}, Ashesh and {Hassanzadeh}, Pedram and {Pasha}, Saba},
        title = "{Predicting clustered weather patterns: A test case for applications of convolutional neural networks to spatio-temporal climate data}",
      journal = {Scientific Reports},
         year = 2020,
        month = jan,
       volume = {10},
          eid = {1317},
        pages = {1317},
          doi = {10.1038/s41598-020-57897-9},
archivePrefix = {arXiv},
       eprint = {1811.04817},
 primaryClass = {physics.ao-ph},
       adsurl = {https://ui.adsabs.harvard.edu/abs/2020NatSR..10.1317C}
}

@ARTICLE{2018arXiv180400684W,
       author = {{Wang}, Bao and {Luo}, Xiyang and {Zhang}, Fangbo and {Yuan}, Baichuan and {Bertozzi}, Andrea L. and {Brantingham}, P. Jeffrey},
        title = "{Graph-Based Deep Modeling and Real Time Forecasting of Sparse Spatio-Temporal Data}",
      journal = {arXiv e-prints},
         year = 2018,
        month = apr,
          eid = {arXiv:1804.00684},
        pages = {arXiv:1804.00684},
archivePrefix = {arXiv},
       eprint = {1804.00684},
 primaryClass = {cs.LG},
       adsurl = {https://ui.adsabs.harvard.edu/abs/2018arXiv180400684W}
}

@ARTICLE{2018arXiv180200386W,
       author = {{Wang}, Leye and {Geng}, Xu and {Ma}, Xiaojuan and {Liu}, Feng and {Yang}, Qiang},
        title = "{Cross-City Transfer Learning for Deep Spatio-Temporal Prediction}",
      journal = {arXiv e-prints},
         year = 2018,
        month = feb,
          eid = {arXiv:1802.00386},
        pages = {arXiv:1802.00386},
archivePrefix = {arXiv},
       eprint = {1802.00386},
 primaryClass = {cs.AI},
       adsurl = {https://ui.adsabs.harvard.edu/abs/2018arXiv180200386W}
}

@ARTICLE{2020arXiv201110616W,
       author = {{Wang}, Rui and {Maddix}, Danielle and {Faloutsos}, Christos and {Wang}, Yuyang and {Yu}, Rose},
        title = "{Bridging Physics-based and Data-driven modeling for Learning Dynamical Systems}",
      journal = {arXiv e-prints},
         year = 2020,
        month = nov,
          eid = {arXiv:2011.10616},
        pages = {arXiv:2011.10616},
archivePrefix = {arXiv},
       eprint = {2011.10616},
 primaryClass = {cs.LG},
       adsurl = {https://ui.adsabs.harvard.edu/abs/2020arXiv201110616W}
}

@inproceedings{10.1145/3394486.3403198,
author = {Wang, Rui and Kashinath, Karthik and Mustafa, Mustafa and Albert, Adrian and Yu, Rose},
title = {Towards Physics-Informed Deep Learning for Turbulent Flow Prediction},
year = {2020},
isbn = {9781450379984},
publisher = {Association for Computing Machinery},
address = {New York, NY, USA},
url = {https://doi.org/10.1145/3394486.3403198},
doi = {10.1145/3394486.3403198},
booktitle = {Proceedings of the 26th ACM SIGKDD International Conference on Knowledge Discovery \& Data Mining},
pages = {1457–1466},
numpages = {10},
location = {Virtual Event, CA, USA},
series = {KDD '20}
}

@phdthesis{Karnowski2012DeepML,
  title={Deep Machine Learning with Spatio-Temporal Inference},
  author={Thomas P. Karnowski},
  year={2012},
  school = {University of Tennessee},
  url = {https://trace.tennessee.edu/utk_graddiss/1315/}
}

@Article{Lin2015,
  author =    {Yun Fei Lin and Suk Kyoung Lee and Pradip Adhikari and Thushani Herath and Steven Lingenfelter and Alexander H. Winney and Wen Li},
  title =     {Note: An improved 3D imaging system for electron-electron coincidence measurements},
  journal =   {Review of Scientific Instruments},
  year =      {2015},
  volume =    {86},
  number =    {9},
  pages =     {096110},
  month =     sep,
  doi =       {10.1063/1.4931684},
  publisher = {{AIP} Publishing},
  url =       {https://doi.org/10.1063/1.4931684}
}

@PhdThesis{Wallauer2011,
  author =   {Robert Wallauer},
  title =    {Untersuchung von Korrelationseffekten in der Doppelphotoemission von normal- und supraleitendem Blei},
  school =   {Universität Frankfurt},
  year =     {2011}
}

@book{MATLAB:2018,
year = {2019},
author = {MATLAB},
title = {9.6.0.1072779 (R2019a)},
publisher = {The MathWorks Inc.},
address = {Natick, Massachusetts}
}

@Article{Becker2012,
  author =    {Becker, Wilhelm and Liu, XiaoJun and Ho, Phay Jo and Eberly, Joseph H.},
  title =     {Theories of photoelectron correlation in laser-driven multiple atomic ionization},
  journal =   {Rev. Mod. Phys.},
  year =      {2012},
  volume =    {84},
  pages =     {1011--1043},
  month =     {Jul},
  doi =       {10.1103/RevModPhys.84.1011},
  issue =     {3},
  numpages =  {0},
  publisher = {American Physical Society},
  url =       {https://link.aps.org/doi/10.1103/RevModPhys.84.1011}
}

@article{Hassan2017,
  doi = {10.1038/nphoton.2017.79},
  url = {https://doi.org/10.1038/nphoton.2017.79},
  year = {2017},
  month = may,
  publisher = {Springer Science and Business Media {LLC}},
  volume = {11},
  number = {7},
  pages = {425--430},
  author = {M. Th. Hassan and J. S. Baskin and B. Liao and A. H. Zewail},
  title = {High-temporal-resolution electron microscopy for imaging ultrafast electron dynamics},
  journal = {Nature Photonics}
}

@article{Shiloh2022,
  doi = {10.1103/physrevlett.128.235301},
  url = {https://doi.org/10.1103/physrevlett.128.235301},
  year = {2022},
  month = jun,
  publisher = {American Physical Society ({APS})},
  volume = {128},
  number = {23},
  author = {R. Shiloh and T. Chlouba and P. Hommelhoff},
  title = {Quantum-Coherent Light-Electron Interaction in a Scanning Electron Microscope},
  journal = {Physical Review Letters}
}

@article{Costa2012,
  doi = {10.1063/1.4770120},
  url = {https://doi.org/10.1063/1.4770120},
  year = {2012},
  month = dec,
  publisher = {{AIP} Publishing},
  volume = {83},
  number = {12},
  pages = {123709},
  author = {G. Da Costa and H. Wang and S. Duguay and A. Bostel and D. Blavette and B. Deconihout},
  title = {Advance in multi-hit detection and quantization in atom probe tomography},
  journal = {Review of Scientific Instruments}
}

@book{van1995python,
  title={Python reference manual},
  author={Van Rossum, Guido and Drake Jr, Fred L},
  year={1995},
  publisher={Centrum voor Wiskunde en Informatica Amsterdam}
}

@article{Albertsson:2018maf,
    author = "Albertsson, Kim and others",
    title = "{Machine Learning in High Energy Physics Community White Paper}",
    eprint = "1807.02876",
    archivePrefix = "arXiv",
    primaryClass = "physics.comp-ph",
    reportNumber = "FERMILAB-PUB-18-318-CD-DI-PPD",
    doi = "10.1088/1742-6596/1085/2/022008",
    journal = "J. Phys. Conf. Ser.",
    volume = "1085",
    number = "2",
    pages = "022008",
    year = "2018"
}

@article{Guest:2018yhq,
    author = "Guest, Dan and Cranmer, Kyle and Whiteson, Daniel",
    title = "{Deep Learning and its Application to LHC Physics}",
    eprint = "1806.11484",
    archivePrefix = "arXiv",
    primaryClass = "hep-ex",
    doi = "10.1146/annurev-nucl-101917-021019",
    journal = "Ann. Rev. Nucl. Part. Sci.",
    volume = "68",
    pages = "161--181",
    year = "2018"
}

@article{Larkoski:2017jix,
    author = "Larkoski, Andrew J. and Moult, Ian and Nachman, Benjamin",
    title = "{Jet Substructure at the Large Hadron Collider: A Review of Recent Advances in Theory and Machine Learning}",
    eprint = "1709.04464",
    archivePrefix = "arXiv",
    primaryClass = "hep-ph",
    doi = "10.1016/j.physrep.2019.11.001",
    journal = "Phys. Rept.",
    volume = "841",
    pages = "1--63",
    year = "2020"
}

@ARTICLE{Radovic2018-no,
  title     = "Machine learning at the energy and intensity frontiers of
               particle physics",
  author    = "Radovic, Alexander and Williams, Mike and Rousseau, David and
               Kagan, Michael and Bonacorsi, Daniele and Himmel, Alexander and
               Aurisano, Adam and Terao, Kazuhiro and Wongjirad, Taritree",
  journal   = "Nature",
  publisher = "Springer Science and Business Media LLC",
  volume    =  560,
  number    =  7716,
  pages     = "41--48",
  month     =  aug,
  year      =  2018,
  language  = "en"
}

@article{Carleo:2019ptp,
    author = "Carleo, Giuseppe and Cirac, Ignacio and Cranmer, Kyle and Daudet, Laurent and Schuld, Maria and Tishby, Naftali and Vogt-Maranto, Leslie and Zdeborov\'a, Lenka",
    title = "{Machine learning and the physical sciences}",
    eprint = "1903.10563",
    archivePrefix = "arXiv",
    primaryClass = "physics.comp-ph",
    doi = "10.1103/RevModPhys.91.045002",
    journal = "Rev. Mod. Phys.",
    volume = "91",
    number = "4",
    pages = "045002",
    year = "2019"
}

@article{Bourilkov:2019yoi,
    author = "Bourilkov, Dimitri",
    title = "{Machine and Deep Learning Applications in Particle Physics}",
    eprint = "1912.08245",
    archivePrefix = "arXiv",
    primaryClass = "physics.data-an",
    doi = "10.1142/S0217751X19300199",
    journal = "Int. J. Mod. Phys. A",
    volume = "34",
    number = "35",
    pages = "1930019",
    year = "2020"
}

@article{Schwartz:2021ftp,
    author = "Schwartz, Matthew D.",
    title = "{Modern Machine Learning and Particle Physics}",
    eprint = "2103.12226",
    archivePrefix = "arXiv",
    primaryClass = "hep-ph",
    doi = "10.1162/99608f92.beeb1183",
    month = "3",
    year = "2021"
}

@article{Karagiorgi:2021ngt,
    author = "Karagiorgi, Georgia and Kasieczka, Gregor and Kravitz, Scott and Nachman, Benjamin and Shih, David",
    title = "{Machine Learning in the Search for New Fundamental Physics}",
    eprint = "2112.03769",
    archivePrefix = "arXiv",
    primaryClass = "hep-ph",
    month = "12",
    year = "2021"
}

@article{Boehnlein:2021eym,
    author = "Boehnlein, Amber and others",
    title = "{Colloquium: Machine learning in nuclear physics}",
    eprint = "2112.02309",
    archivePrefix = "arXiv",
    primaryClass = "nucl-th",
    doi = "10.1103/RevModPhys.94.031003",
    journal = "Rev. Mod. Phys.",
    volume = "94",
    number = "3",
    pages = "031003",
    year = "2022"
}

@phdthesis{bauer:Thesis,
  author       = {Tobias Bauer}, 
  title        = {Koinzidente Photoelektronenspektroskopie an Kuprat-Hochtemperatursupraleitern},
  school       = {Goethe-Universit\"at Frankfurt am Main},
  year         = 2015,
}

@Article{Jagutzki2002_2,
  author =  {Jagutzki, O. and Cerezo, A. and Czasch, A. and Dorner, R. and Hattas, M. and Min Huang and Mergel, V. and Spillmann, U. and Ullmann-Pfleger, K. and Weber, T. and Schmidt-Bocking, H. and Smith, G.D.W.},
  title =   {Multiple hit readout of a microchannel plate detector with a three-layer delay-line anode},
  journal = {IEEE Transactions on Nuclear Science},
  year =    {2002},
  volume =  {49},
  number =  {5},
  pages =   {2477-2483},
  doi =     {10.1109/TNS.2002.803889}
}

@Article{Basnayake2021,
  author =    {Basnayake, Gihan and Fernando, Shanilke and Lee, Suk Kyoung and Debrah, Duke A. and Stewart, Gabriel A. and Li, Wen},
  title =     {The lack of electron momentum correlation in strong-field triple ionisation of molecules},
  journal =   {Molecular Physics},
  year =      {2021},
  volume =    {120},
  number =    {1–2},
  month =     may,
  doi =       {10.1080/00268976.2021.1931722},
  issn =      {1362-3028},
  publisher = {Informa UK Limited},
  url =       {http://dx.doi.org/10.1080/00268976.2021.1931722}
}

@Article{Meier2023,
  author =    {Stefan Meier and Jonas Heimerl and Peter Hommelhoff},
  title =     {Few-electron correlations after ultrafast photoemission from nanometric needle tips},
  journal =   {Nature Physics},
  year =      {2023},
  volume =    {19},
  pages =     {1402-1409},
  month =     jun,
  doi =       {10.1038/s41567-023-02059-7},
  publisher = {Springer Science and Business Media {LLC}},
  url =       {https://doi.org/10.1038/s41567-023-02059-7}
}

@article{Haindl2023,
  title = {Coulomb-correlated electron number states in a transmission electron microscope beam},
  volume = {19},
  ISSN = {1745-2481},
  url = {http://dx.doi.org/10.1038/s41567-023-02067-7},
  DOI = {10.1038/s41567-023-02067-7},
  number = {10},
  journal = {Nature Physics},
  publisher = {Springer Science and Business Media LLC},
  author = {Haindl,  Rudolf and Feist,  Armin and Domr\"{o}se,  Till and M\"{o}ller,  Marcel and Gaida,  John H. and Yalunin,  Sergey V. and Ropers,  Claus},
  year = {2023},
  month = jun,
  pages = {1410–1417}
}
\end{document}